%% file: ubicomp_main.tex
\documentclass[acmlarge,screen,nonacm]{acmart}

\usepackage{graphicx}
\usepackage{subcaption}
\usepackage{xspace}
\AtBeginDocument{%
  \providecommand\BibTeX{{%
    \normalfont B\kern-0.5em{\scshape i\kern-0.25em b}\kern-0.8em\TeX}}}

\setcopyright{none}
\renewcommand\footnotetextcopyrightpermission[1]{} 
\usepackage{colortbl}
\usepackage[british]{babel}
\usepackage{hhline}
\usepackage{multirow}
\usepackage{makecell}

\begin{document}

\newcommand{\sysname}{IoTProsector\xspace}

\newcommand{\haojian}[1]{\textcolor{violet}{#1}}


\title{On the Feasibility of Reasoning about the Internal States of Blackbox IoT Devices Using Side-Channel Information}


\input{ubicomp_parts/p0-abstract.tex}

\author{Wei Sun}
\authornote{Both authors contributed equally}
\affiliation{%
  \institution{University of California San Diego}
  \country{USA}}
\email{w5sun@ucsd.edu}

\author{Yuwei Xiao}
\authornotemark[1]
\affiliation{%
  \institution{University of California San Diego}
  \country{USA}}
\email{yux075@ucsd.edu}

\author{Haojian Jin}
\affiliation{%
  \institution{University of California San Diego}
  \country{USA}}
\email{haojian@ucsd.edu}

\author{Dinesh Bharadia}
\affiliation{%
  \institution{University of California San Diego}
  \country{USA}}
\email{dineshb@ucsd.edu}

\begin{teaserfigure}
\includegraphics[width=\textwidth]{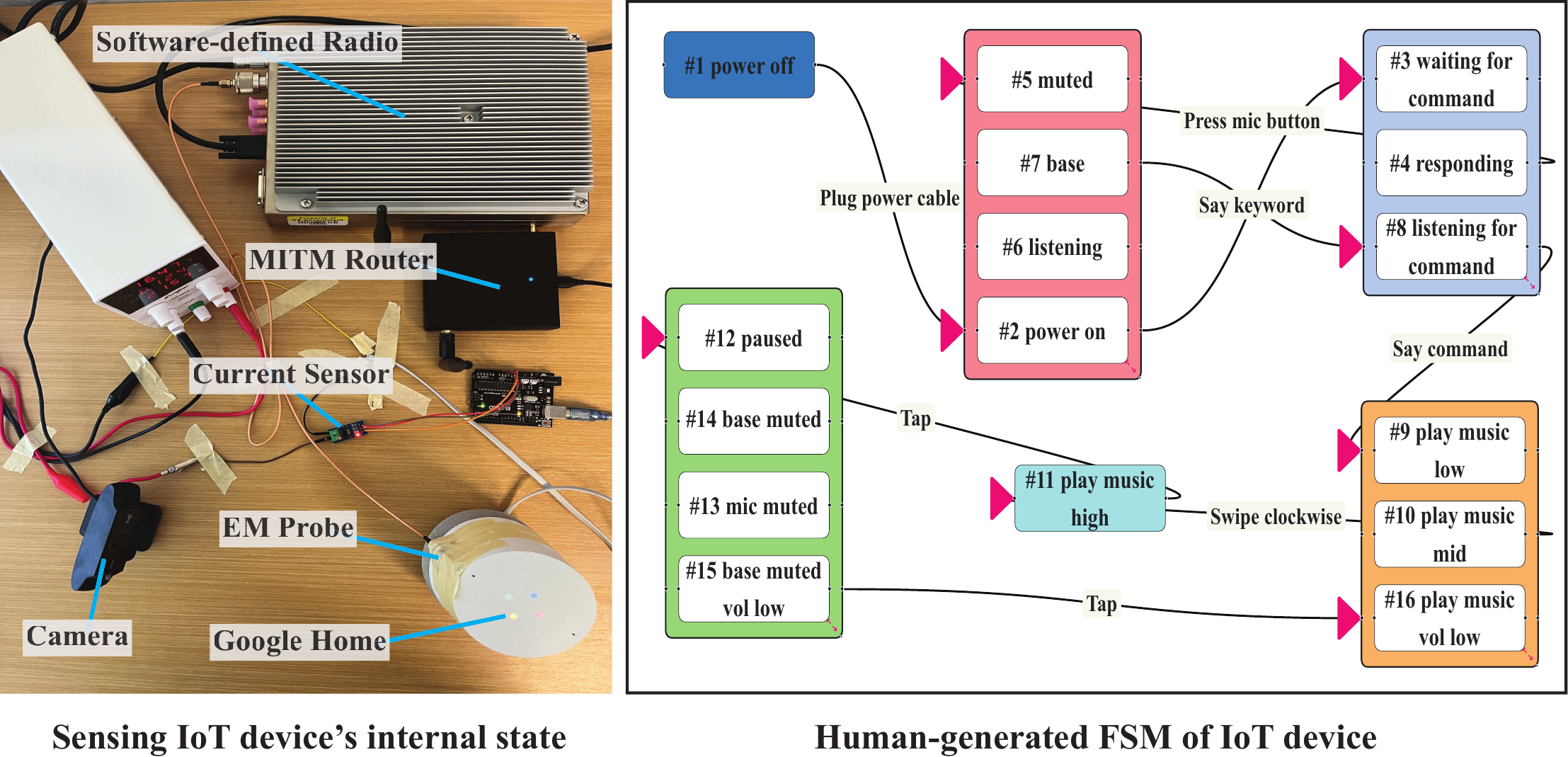}
\caption{\sysname helps users reason the internal states of black-box IoT devices (e.g., Google Home smart speaker) by sensing side-channel information (e.g., power, emanations, and network traffic) and incorporating human understandings with a user interface design, resulting in the human-generated finite state machine (FSM) of the IoT device.  
} 
\Description{Explain the general flow of the system.} 
\label{fig:teaser}
\end{teaserfigure}


\maketitle
\pagestyle{plain}

\input{ubicomp_parts/p1-intro}

\input{ubicomp_parts/p3-overview}
\input{ubicomp_parts/p4-principle}

\input{ubicomp_parts/p5-implementation}

\input{ubicomp_parts/p6-results}
\input{ubicomp_parts/p7-related}

\input{ubicomp_parts/p8-conclusion}
\input{ubicomp_parts/p10-ackknowledgement}

\bibliographystyle{ACM-Reference-Format}
\bibliography{sample-base}

\input{ubicomp_parts/p9-appendix}

\end{document}

%% file: ubicomp_parts/p0-abstract.tex
\begin{abstract}


Internet of Things (IoT) devices are typically designed to function in a secure, closed environment, making it difficult for users to comprehend devices' behaviors.
This paper shows that a user can leverage side-channel information to reason fine-grained internal states of black box IoT devices. 
The key enablers for our design are a multi-model sensing technique that fuses power consumption, network traffic, and radio emanations and an annotation interface that helps users form mental models of a black box IoT system. 
We built a prototype of our design and evaluated the prototype with open-source IoT devices and black-box commercial devices. 
Our experiments show a false positive rate of 1.44\% for open-source IoT devices' state probing, and our participants take an average of 19.8 minutes to reason the internal states of black-box IoT devices. 

\end{abstract}

\ccsdesc[500]{Human-centered computing~Ubiquitous and mobile devices}
\ccsdesc[500]{Human-centered computing~Graphical user interfaces}
\keywords{Side-Channel sensing, sense-making, mental model, system images}

%% file: ubicomp_parts/p1-intro.tex
\section{Introduction}
\begin{figure}
  \centering
  \includegraphics[width=\linewidth]{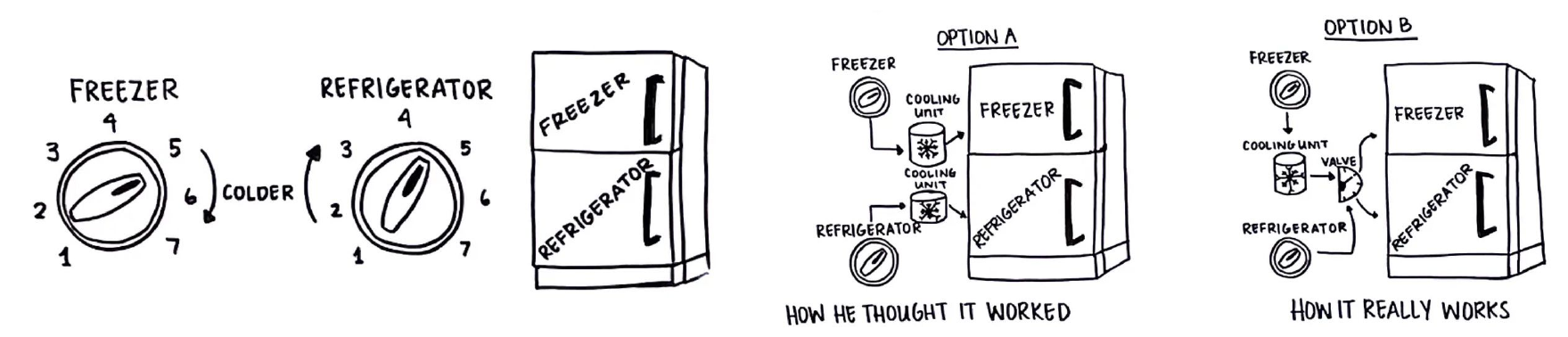}
    \caption{Don Norman owned a two-compartment refrigerator with two controls: a freezer and a refrigerator. To his surprise, when he tried to make the freezer colder, it also made the refrigerator colder - even when he didn't change the refrigerator dial. Indeed, his mental model (option A) of how the cooling unit worked differed from how it actually worked (option B): there was only one cooling unit with a valve that dispersed the air to each compartment, instead of two respective cooling units. We adapt the image from~\cite{March14M81:online}.}
    \label{fig:refrigerator}
\end{figure}

In "The Design of Everyday Things~\cite{norman1988psychology}", Don Norman used an example of a refrigerator (Fig.~\ref{fig:refrigerator}) to illustrate that users often have wrong mental models of how products work. As described by Norman, users generate a mental model for how their interactions affect the system and how the system affects them through reading product descriptions and observing system behaviors. However, unlike the refrigerator example, many low-level system behaviors in IoT devices are intangible. As a result, users have various questions regarding the internal states of these devices, such as: "Do Amazon Echo devices really stop listening, when a user presses the mute button~\cite{EchoMute41:online}?", "Why do cameras stop recording after 30 minutes~\cite{AskHNWhy9:online}?"

A number of HCI studies have tried to understand the mental models of smart home users. For example, Clark et al.~\cite{clark2017devices} found that different smart home abstractions have significant priming effects on users' mental models. Blase et al. examined whether trigger-action programming (e.g., IFTTT programs) captures smart home behaviors that users actually desire~\cite{ur2014practical}. 
In contrast to these studies, our work focuses on building a new tool to support users in forming mental models of IoT devices. Particularly, we consider the side-channel information as a new channel for understanding intangible system behaviors and designing interactions accordingly. 

This paper presents \sysname, a system that leverages multimodal side-channel information (i.e., power consumption, network traffic, and electromagnetic emanations) to help users understand the fine-grained internal states of IoT devices. 
\sysname first records how an experimenter interacts with a target IoT device and captures the side channel information generated along the process. 
By clustering the sensor data distributions and analyzing the temporal transitions, \sysname then derives a finite state machine and helps experimenters align the IoT device's internal states with their mental models (Fig.~\ref{fig:teaser}).

\sysname\ has two key enablers. The first is a multi-model sensing technique that correlates the side-channel information with the IoT device's internal states. When the IoT device stays in different states, its power consumption and generated network traffic differ. For example, the Nest Cam's power consumption remained almost identical when in "indicator-off" mode (340 mA) as when fully operational (370 mA)~\cite{NestCamW80:online}. This slight reduction correlates with the disabling of the LED power light, given that LEDs typically draw 10-20mA. Further, the electromagnetic emanations are amplitude-modulated clock signals caused by the computation activities on the IoT device, which can exhibit periodic spikes in the frequency domain, and their power spectral densities correlate with the IoT device's internal states. Therefore, we can fuse this side-channel information and leverage machine learning models (e.g., k-means) to probe the IoT device's internal states.

Second, we develop an annotation user interface to bridge the gap between low-level states and high-level mental models. For example, a smart speaker that plays music at different volumes may have distinct side-channel information, but these states may share similar cognitive meanings. \sysname\ introduces a four-step workflow (explore-model-collage-verify) to guide users in forming mental models using side-channel information. 
\sysname\ begins by allowing users to blindly interact with the target IoT devices and collecting side-channel information. \sysname\ then identifies unique states using the side-channel information and asks users to merge redundant states. Finally, users verify the correctness of the mental model by interacting with the IoT devices step-by-step.

We built a prototype of our design and evaluated the prototype with open-source IoT
devices (Google AIY Voice kits and Vision kits) and black-box commercial devices (Google Home). Our experiments show a false positive rate of 1.44\% for open-source IoT devices’
state probing, and our participants take an average of 19.8 minutes to reason the internal states of black-box IoT devices. Notably, it was observed that users' ultimate mental models, shaped by newly acquired side-channel information, continue to differ. This variation may be partly due to the lasting impact of their initial mental models, which affect the way users revise their understanding~\cite{bibby1996instruction}.

Our main contribution is (1) an IoT probing system that can probe the internal states of the black box IoT devices using the side-channel information; (2) a four-step workflow that can connect the low-level system states to the high-level states in users' mental model; (3) Our experimental results demonstrate the precision of \sysname on probing IoT internal states. Our user study further confirms its capability to assist users in annotating both precise and semantically meaningful IoT states.

%% file: ubicomp_parts/p3-overview.tex
\begin{figure}
  \centering
  \includegraphics[width=\linewidth]{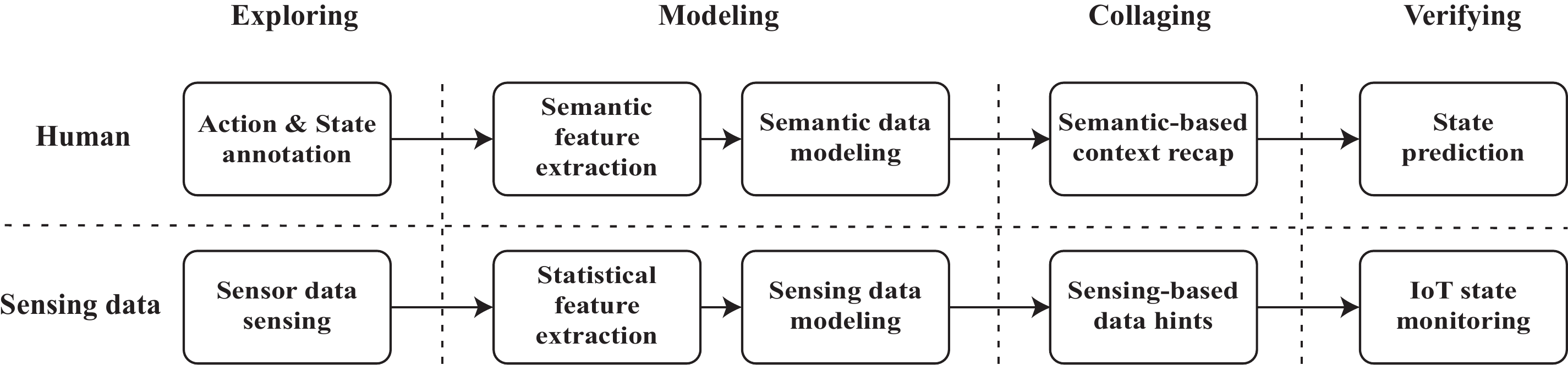}
    \caption{\sysname's workflow consisting of state exploration, sensing, modeling, collaging, and step-wise verification fuses the sensing model and human mental model for accurate and semantic IoT internal state probing.}
    \label{fig:workflow}
\end{figure}

\section{Overview}
\label{sec:overview}
In this section, we present the overview system design of \sysname, which mainly consists of two components. 

\vspace{0.1cm}\noindent\textbf{Sensing.}  We develop a multi-modal sensing technique that leverages and fuses power consumption, network traffic, and emanations to detect the internal states of IoT devices. Specifically, we extract statistical features from the sensing data and apply the TSNE algorithm~\cite{wattenberg2016use} to identify the significant features. 
We then use these features to cluster the IoT device's internal states through unsupervised machine-learning algorithms.

\vspace{0.1cm}\noindent\textbf{Annotation.} Furthermore, we design an annotation user interface to bridge the gap between users' cognitive understanding and low-level machine states derived from side-channel information. We illustrate this process shown in Fig.~\ref{fig:workflow} as follows:  

\begin{itemize}
    \item \textbf{Exploring.}
    A user may first blindly explore the possible internal states of the given IoT device by interacting with the IoT device based on the instructions illustrated in \sysname's graphical user interface. During the interaction, the sensing data generated by the IoT device are collected, and the interaction actions, together with the IoT device's behaviors, are video-tapped to facilitate the later sense-making process. 
    \item \textbf{Modeling.}
    In this process, the sensing data collected during exploration are utilized to extract statistical features and then develop the sensing model. In addition, the transition events between the IoT device's states are employed to merge states, thereby reducing redundancy.
    \item \textbf{Collaging.} 
    To further accurately and semantically probe the IoT device's internal states, a visual and interactive display showcasing the relationship between the sensing model and the human mental model, along with the contextual information is provided to enhance the user's understanding and aid in making a collage of the IoT states. By utilizing this information, the user is expected to generate a sense-making FSM that effectively characterizes the IoT device's internal states.
    \item \textbf{Verifying.}
    After the internal states of the IoT device have been annotated and characterized, the user can verify them using the generated finite-state machine. Specifically, a machine learning classifier is trained based on the collaging results. Then, as the user interacts with the IoT device, the generated sensing data can be classified into one of these internal states. This process enables the user to monitor and analyze the states of the IoT device in real-time. 
\end{itemize}

%% file: ubicomp_parts/p4-principle.tex

\section{\sysname: Sensing}
\label{sec:iot:sensing}

\subsection{Side-channel Information Characterization}

\subsubsection{Power consumption and network traffic} 

Every IoT device drains energy either from the power line or battery and the amount of the energy drained by the IoT device highly depends on the IoT device's state. Even though the network traffic data is usually encrypted, we can still use the network traffic pattern as the side-channel information to probe the IoT devices' internal states. This is because the network traffic introduced by the IoT device highly depends on its state.

\subsubsection{Emanations}

Every IoT device introduces electromagnetic emanations, which are amplitude-modulated clock signals. To identify these emanations from the IoT devices, we can perform a Fast Fourier Transform (FFT) on the collected emanation signals. As a result, the FFT peaks are equally separated in the frequency domain. Then, we use the power spectral density of these FFT peaks as our emanation features for IoT state probing. 

\begin{figure}
  \centering
  \includegraphics[width=0.5\linewidth]{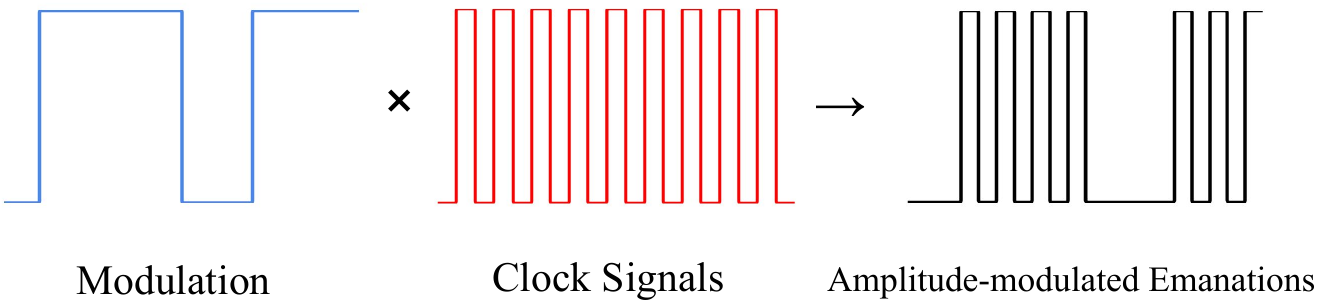}
    \caption{The clock signals are amplitude-modulated by the computation activity on the IoT device, resulting in the amplitude-modulated clock signals that are termed emanations.}
    \label{fig:amp:clock}
\end{figure}

\vspace{0.1cm}\noindent\textbf{Connecting emanations to internal fine-grained states.} Every IoT device has its own clock for synchronization purposes during the computation. The clock signals (i.e., electromagnetic waves) can emit over the air directly, or go through the electronic components on the IoT device and emit over the air afterwards~\cite{sehatbakhsh2019emma}. These clock signals can be further modulated by the computation activities on the IoT devices resulting in the amplitude-modulated clock signals as shown in Fig.~\ref{fig:amp:clock}. Intuitively, when there is a computation activity, there is clock signal leakage. Otherwise, there is no clock signal leakage. So, these clock signals are amplitude-modulated. We term these amplitude-modulated clock signals as emanations.

Since the emanations are amplitude-modulated clock signals, these emanations can carry sensitive data information about the IoT device's internal states. Specifically, IoT devices' states depend on the computation activities conducted on them, which can be revealed by the emanations. So, we can predict the IoT devices' internal states based on the received emanations. In the time domain, the emanations are squared waves as they are amplitude-modulated clock signals, which can exhibit periodic spikes in the frequency domain.

\begin{figure}
\centering
\begin{minipage}{0.5\textwidth}
\captionsetup{width=0.96\textwidth}
\centering
    \includegraphics[width=\linewidth]{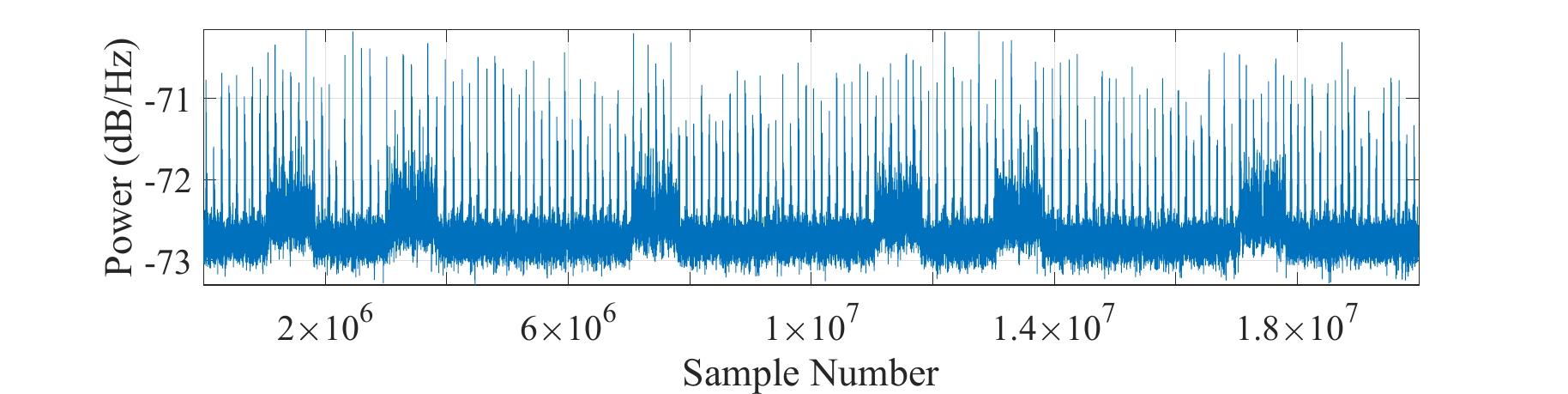}
 \caption{Time-domain emanation signals from Google Home smart speaker exhibit on-off property, as they are amplitude-modulated clock signals.}
    \label{fig:em:time}
\end{minipage}%
\begin{minipage}{0.5\textwidth}
\captionsetup{width=0.96\textwidth}
\centering
    \includegraphics[width=\linewidth]{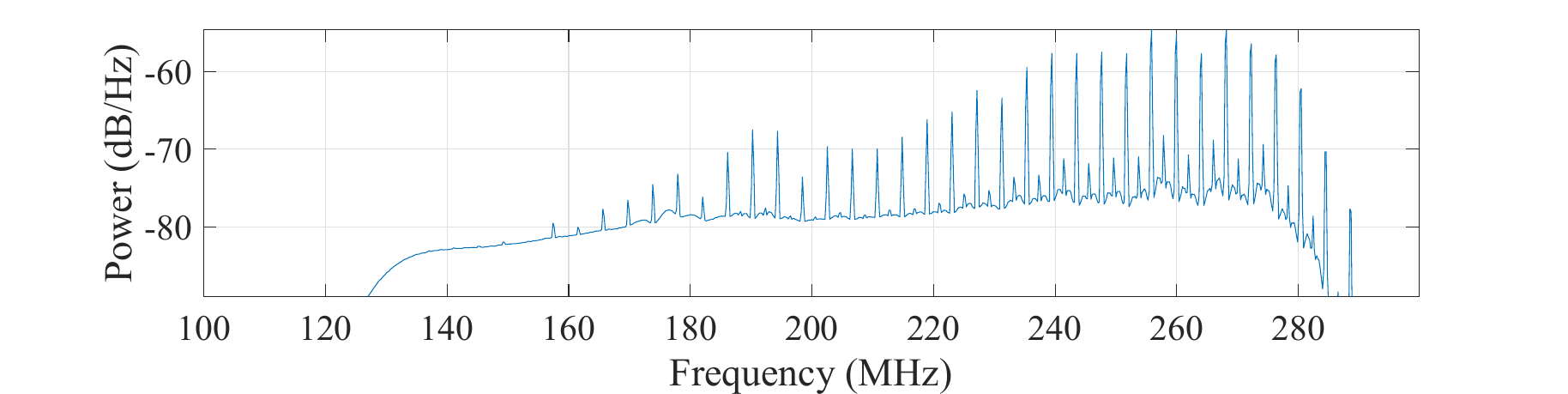}
    \caption{Frequency-domain emanation signals from Google Home smart speaker exhibit periodic spikes spread across the spectrum, which can be leveraged to sense the IoT device's internal states using the spike's power spectral density.}
    \label{fig:em:frequency}
\end{minipage}
\end{figure}

\vspace{0.1cm}\noindent\textbf{Time-domain emanations.} Since the emanations are amplitude-modulated clock signals, they will become the squared waves in the time domain. Therefore, the time-domain emanations present the on-off property. The ideal squared wave using Fourier expansion with a cycle frequency of $f$ over time $t$ can be represented as follows:
\begin{equation}
    x(t)=\frac{4}{\pi}\sum_{k=1}^{\infty }\frac{sin(2\pi(2k-1)ft)}{2k-1}
\end{equation}
To demonstrate the on-off property of the time-domain emanations, Fig.~\ref{fig:em:time} illustrates the emanations of the Google Home smart speaker at the frequency band between 100MHz and 300MHz, which exhibit the on-off property over time. This is because the emanations are amplitude-modulated clock signals.

\vspace{0.1cm}\noindent\textbf{Frequency-domain emanations.} The amplitude-modulated clock signals (i.e., emanations) in the time domain are the squared waves. When we do the Fast Fourier Transform (FFT) on the time-domain emanations, we will have frequency-domain emanations, which will have one fundamental harmonic and other multiple harmonics with decreasing power spectral densities. Specifically, the frequency-domain emanations can be represented as follows:
\begin{equation}
    x(f)=\sum_{k=-\infty}^{\infty}\frac{2sin(2\pi kf_0T)}{k}\delta (f-kf_0)
\end{equation}
where $f_0=\frac{1}{T}$ is the frequency of the fundamental harmonic, and $\delta(f-kf_0)$ indicates the harmonic component at frequency of $kf_0$ with amplitude of $\frac{2sin(2\pi kf_0T)}{k}$. As we can see, the squared wave consists of an infinite number of sine wave components. Moreover, since the emanations are amplitude-modulated, they will spread over the spectrum. Said differently, each sine wave component acts as a different carrier for the modulation signals. Fig.~\ref{fig:em:frequency} showcases the frequency-domain emanations of the Google Home smart speaker at the frequency band between 100MHz and 300MHz. As we can see, each peak in the plot represents the intermediate frequency of the harmonics, which are equally separated as the emanations are the amplitude-modulated clock signals.


The emanations propagate inside and outside the IoT device's circuit. Suppose the leaked emanations at the frequency of $f_l$ inside the circuit, which can amplitude modulate the clock signals at the frequency of $f_c$. The RF transceivers further shape the emanations at the carrier frequency of $f_{carrier}$, resulting in the emanations emitted over the air through the transceiver's antennas. Therefore, the frequencies of the emanations received over the air can be as follows:
\begin{equation}
    f_r=p\cdot f_{carrier} + q\cdot f_c + r\cdot f_l
\end{equation}
where $p, q$, and $r$ are integers due to the mixing and amplification of the emanations. Note that not all of these multiples are applied, which is highly dependent on the hardware architecture and components (e.g., filters) of the circuit. Some IoT devices may even not have RF transceivers, thereby the emanations are emitted through the data lines (i.e., acting as antennas) in the circuit. As we can see, the $f_r$ of the received emanations can be highly dependent on $f_l$ due to the computing activities on the IoT device, which is dependent on the state of the device. Therefore, we can leverage the frequency-domain analysis of the emanations to infer internal IoT states. 

\begin{figure*}
\centering
\begin{minipage}{0.33\textwidth}
\captionsetup{width=0.96\textwidth}
\centering
    \includegraphics[width=\linewidth]{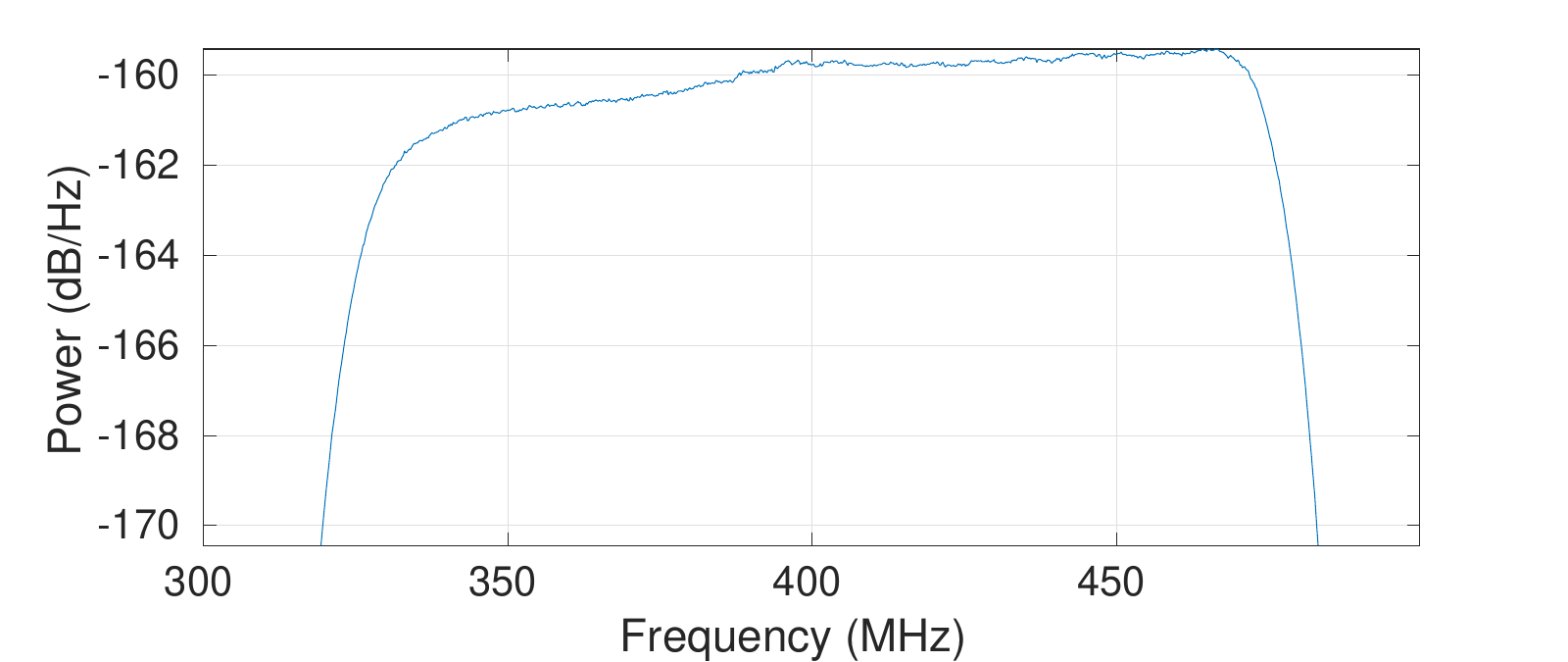}
 \caption{FFT of IQ samples, when the Amazon Echo Dot is power-off.}
    \label{fig:dot:off}
\end{minipage}%
\begin{minipage}{0.33\textwidth}
\captionsetup{width=0.96\textwidth}
\centering
    \includegraphics[width=\linewidth]{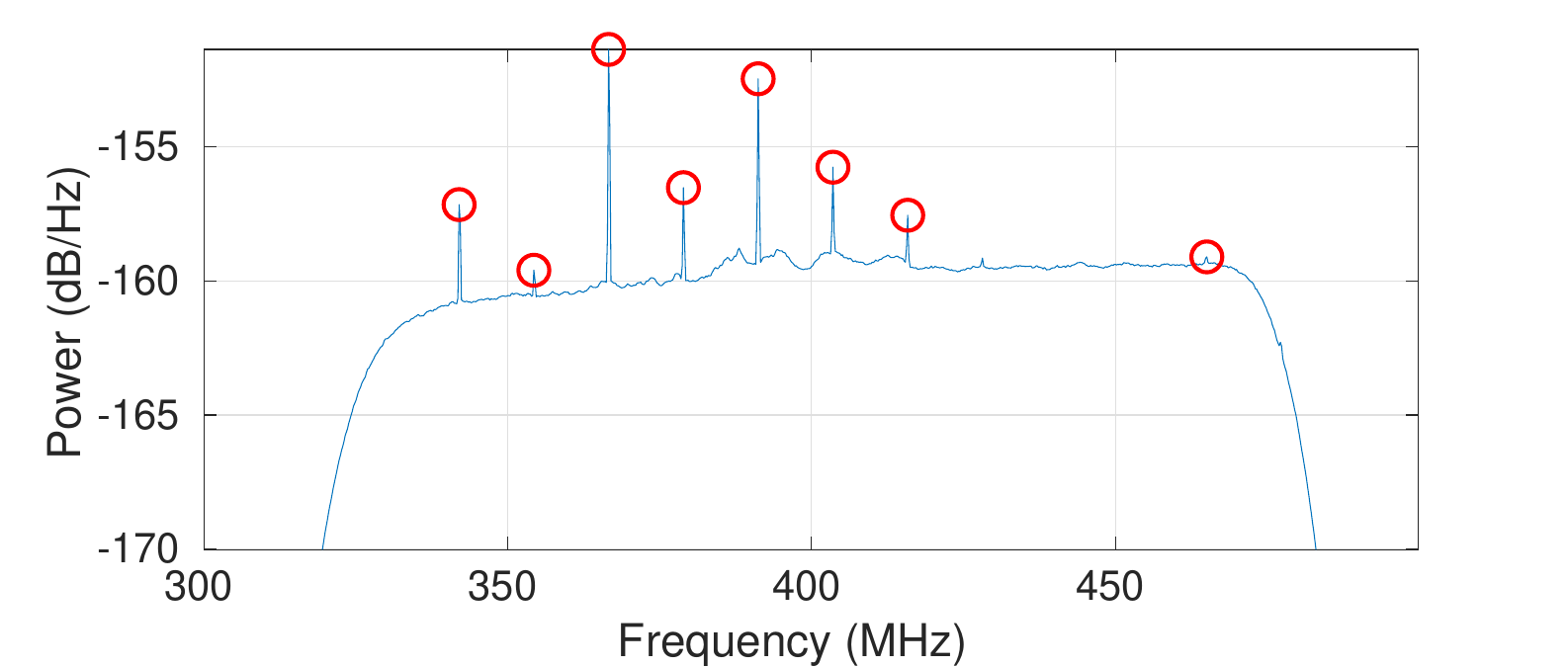}
    \caption{FFT of IQ samples, when the Amazon Echo Dot is power-on.}
    \label{fig:dot:on}
\end{minipage}%
\begin{minipage}{0.33\textwidth}
\captionsetup{width=0.96\textwidth}
\centering
   \includegraphics[width=\linewidth]{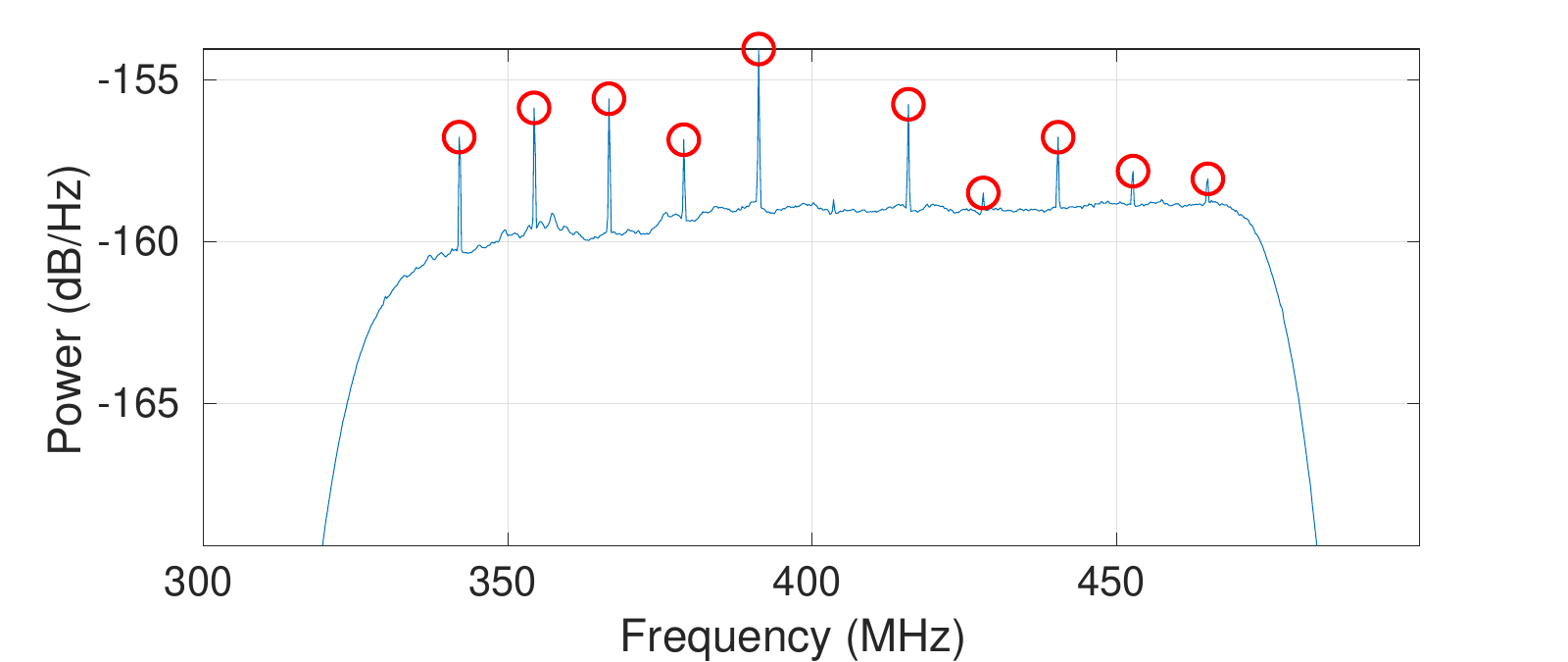} 
    \caption{FFT of IQ samples, when we interact with the Amazon Echo Dot.}
    \label{fig:dot:interaction}
\end{minipage}
\end{figure*}

\begin{figure*}
\centering
\begin{minipage}{0.33\textwidth}
\captionsetup{width=0.96\textwidth}
\centering
    \includegraphics[width=\linewidth]{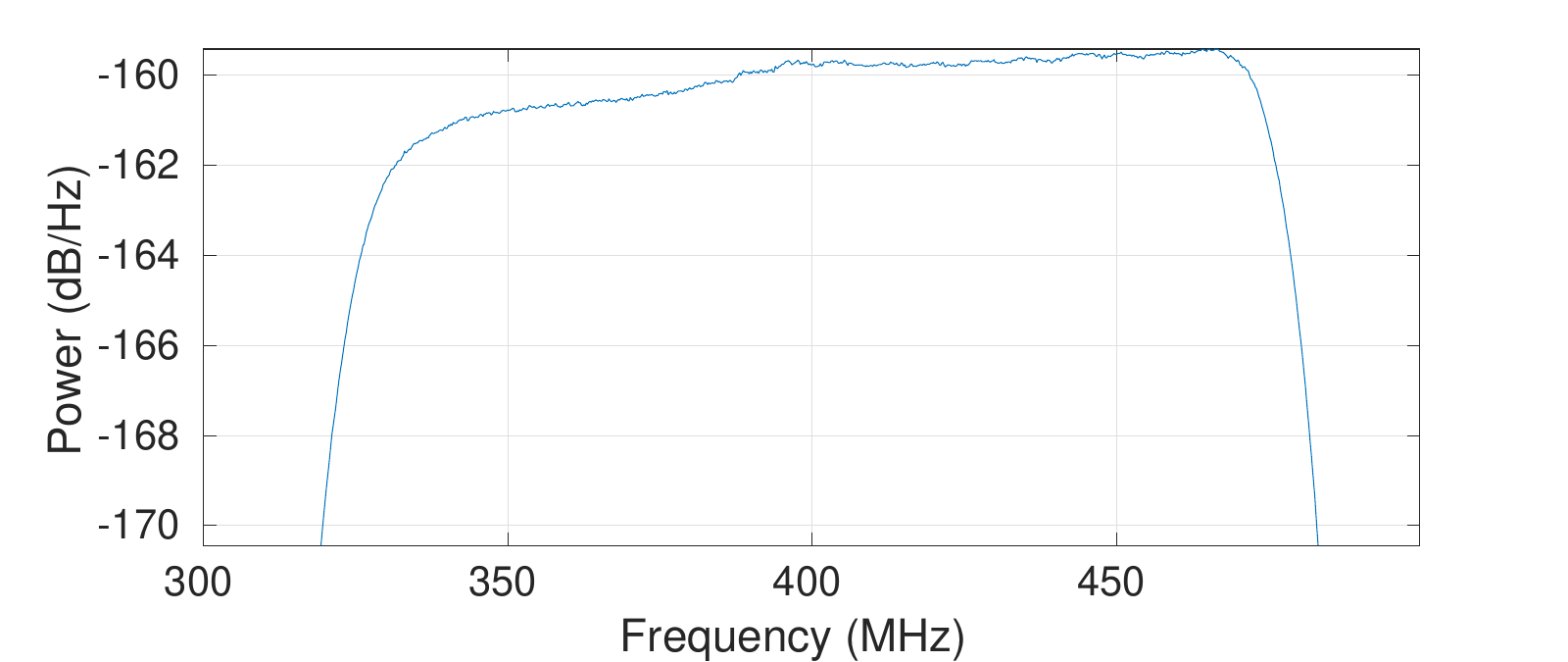}
 \caption{FFT of IQ samples, when the Google Home smart speaker is power-off.}
    \label{fig:google:off}
\end{minipage}%
\begin{minipage}{0.33\textwidth}
\captionsetup{width=0.96\textwidth}
\centering
    \includegraphics[width=\linewidth]{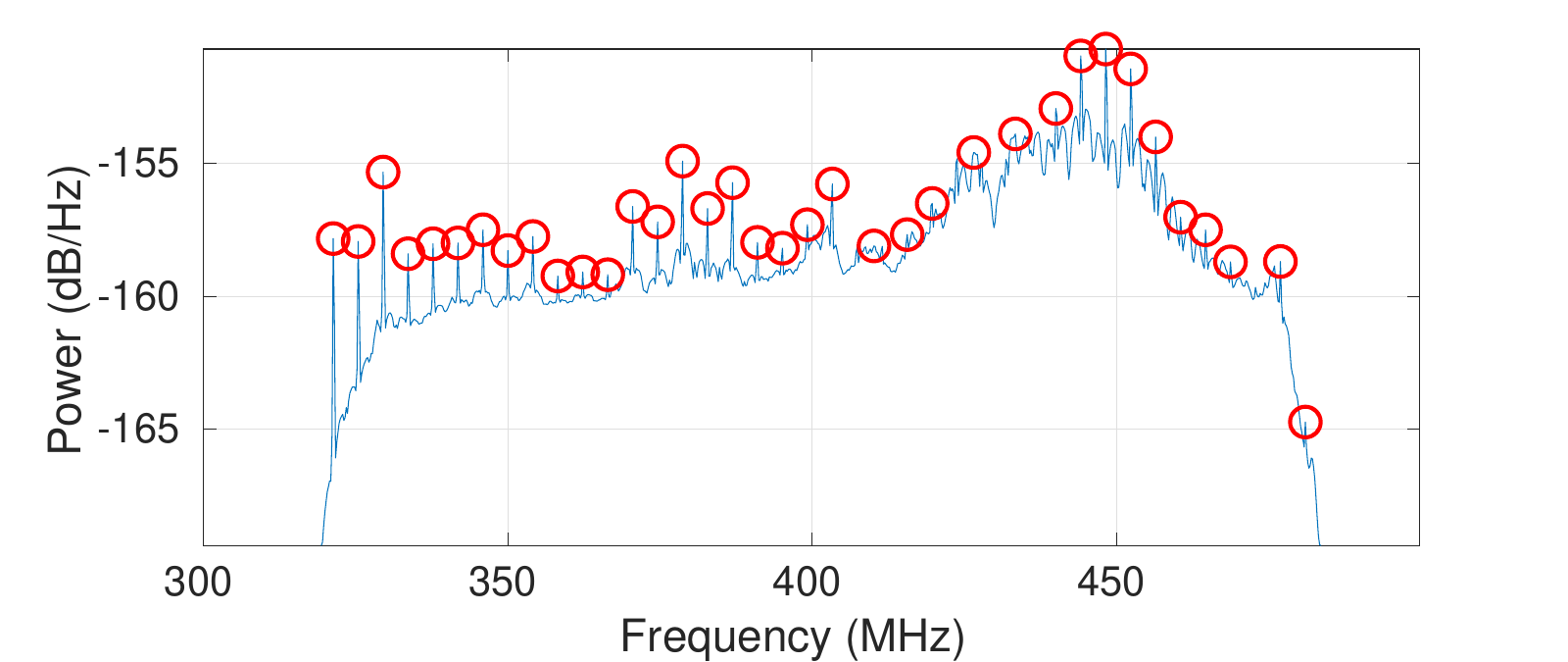}
    \caption{FFT of IQ samples, when the Google Home smart speaker is power-on.}
    \label{fig:google:on}
\end{minipage}%
\begin{minipage}{0.33\textwidth}
\captionsetup{width=0.96\textwidth}
\centering
   \includegraphics[width=\linewidth]{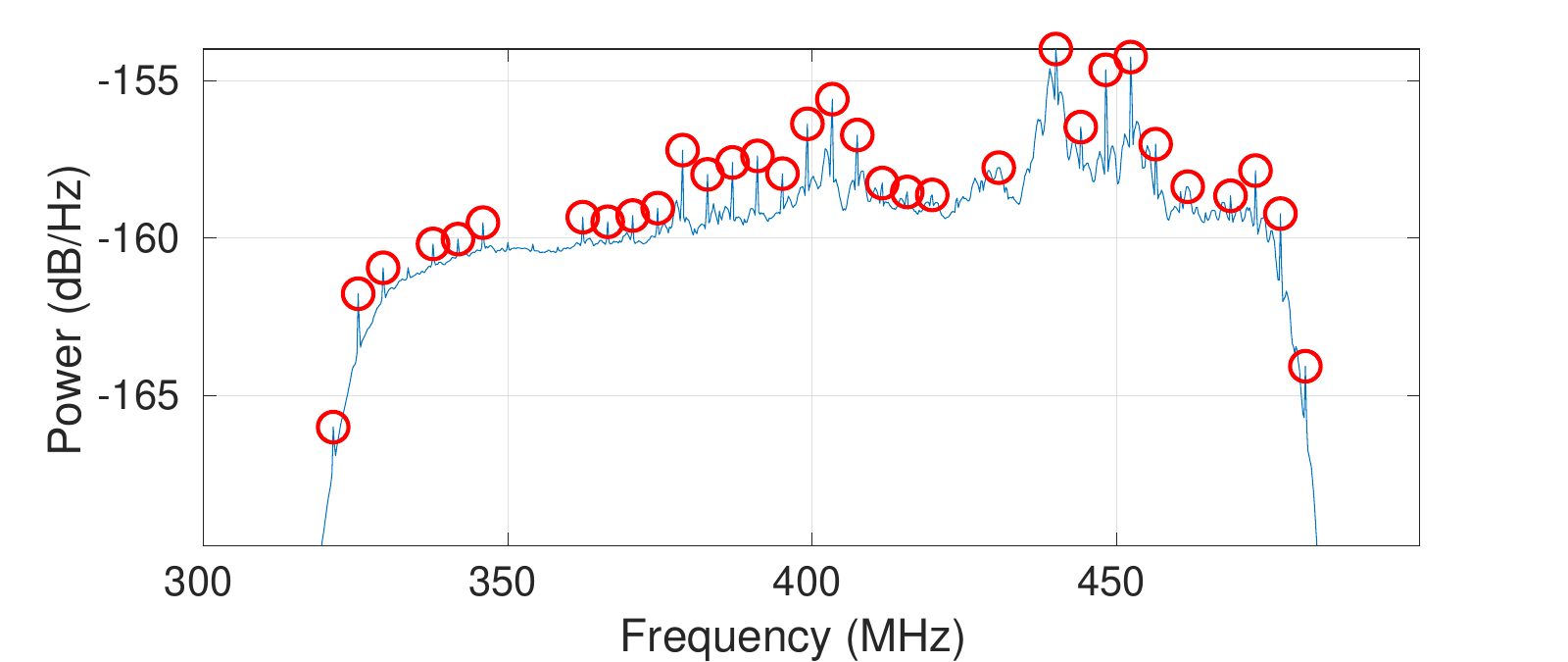} 
    \caption{FFT of IQ samples, when we interact with the Google Home smart speaker.}
    \label{fig:google:interaction}
\end{minipage}
\end{figure*}

\subsection{Feasibility Study}
To demonstrate the feasibility of using the IoT device's emanations for internal IoT state detection, we measure the emanations from the Amazon Echo Dot and Google Home smart speaker. Specifically, Fig.~\ref{fig:dot:off}, Fig.~\ref{fig:dot:on}, and Fig.~\ref{fig:dot:interaction} show the frequency-domain emanations from the Amazon Echo Dot, when it is power-off, power-on, and interacting with people respectively. Fig.~\ref{fig:google:off}, Fig.~\ref{fig:google:on}, and Fig.~\ref{fig:google:interaction} shows the frequency-domain emanations from the Google Home smart speaker, when it is power-off, power-on, and interacting with people respectively. The red circle indicates the detected FFT peaks on the frequency-domain emanations. As we can see, the Google Home smart speaker and Amazon Echo Dot have different emanation patterns when they are in different states. Moreover, the Google Home smart speaker and Amazon Echo Dot will exhibit different emanation patterns, even though they are in the same state. This is because different IoT devices will have different emanations due to the different hardware architectures. 
\begin{table}[t]
\centering
\small
\begin{tabular}{|c|c|}
\hline
\textbf{Statistical features} & \textbf{Description}                \\ \hline
MAV                  & mean absolute value        \\ \hline
VAR                  & variance                   \\ \hline
RMS                  & root mean square           \\ \hline
Std                  & standard deviation         \\ \hline
MAD                  & median absolute deviation  \\ \hline
Skewness             & asymmetry of the data distribution \\ \hline
Kurtosis             & shape of the data distribution                   \\ \hline
IQR                  & interquartile range         \\ \hline
Energy               & average sum of the squares \\ \hline
\end{tabular}
\caption{Statistical features for time-series power consumption, network traffic, and electromagnetic emanations over frequencies.}
\label{table:features}
\end{table}

\subsection{Machine Learning-based IoT State Probing}
Using the side-channel information collected over time during exploration, we compute features from this data as the input for the sensing model to probe the IoT device’s internal states. As shown in Table~\ref{table:features}, we list nine statistical features that we can extract from this side-channel information. Specifically, we measure the time-series network throughput for network traffic data collection, which will be used to derive the statistical features for network traffic side-channel information. Similarly, we measure the time-series power consumption of the IoT device in each state and derive the statistical features for power consumption side-channel information. However, for the emanation side-channel information, we first collect the time-domain IQ samples and further conduct Fast Fourier Transform (FFT) to obtain the frequency domain signals. The intuition is that the IoT device's states are related to the power density of the spikes presented at the frequency domain IQ samples, which is illustrated in the above section. Then, we use the power density of spikes presented in frequency domain signals as the series of data streams to derive the statistical features. After we obtain the statistical features from all the side-channel information, we concatenate them to formulate a vector. Since this three side-channel information could play a different role in IoT state prediction, it's important to only extract some important features to efficiently train the machine learning models for IoT state prediction. Therefore, we concatenate these features and use the two most important features determined by the t-SNE algorithm~\cite{van2008visualizing} as the input of unsupervised classification models (i.e., k-means, DBSCAN, and GMM) for IoT state probing.

\section{\sysname: Annotation}
\label{sec:iot:debugging}

\begin{figure*}[t!]
    \centering
    \begin{subfigure}[t]{0.5\textwidth}
        \centering
        \captionsetup{width=0.96\textwidth}
        \includegraphics[width=\textwidth]{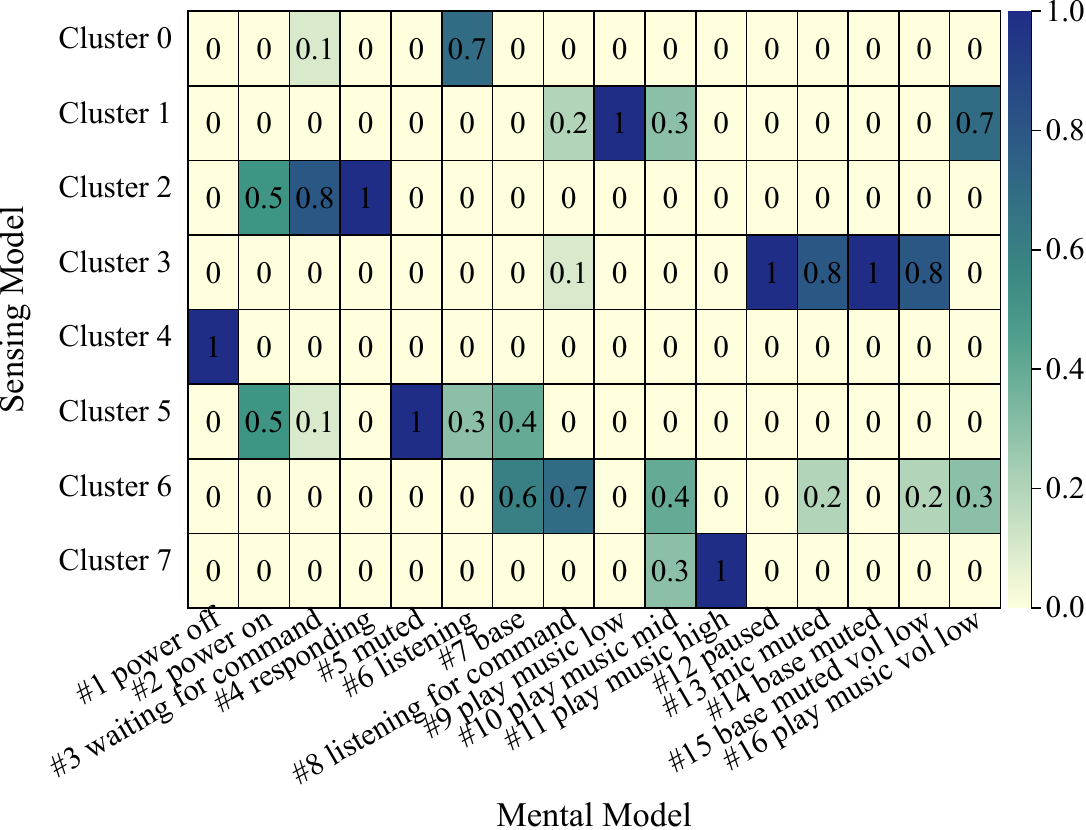}
        \caption{Correlation matrix of the IoT states derived from the sensing model and human mental model, where each element in the correlation matrix indicates the percentage of the human-annotated states on the cluster generated by the sensing model.}
        \label{fig:cm:collage:1}
    \end{subfigure}%
    ~ 
    \begin{subfigure}[t]{0.5\textwidth}
        \centering
        \captionsetup{width=0.96\textwidth}
        \includegraphics[width=\textwidth]{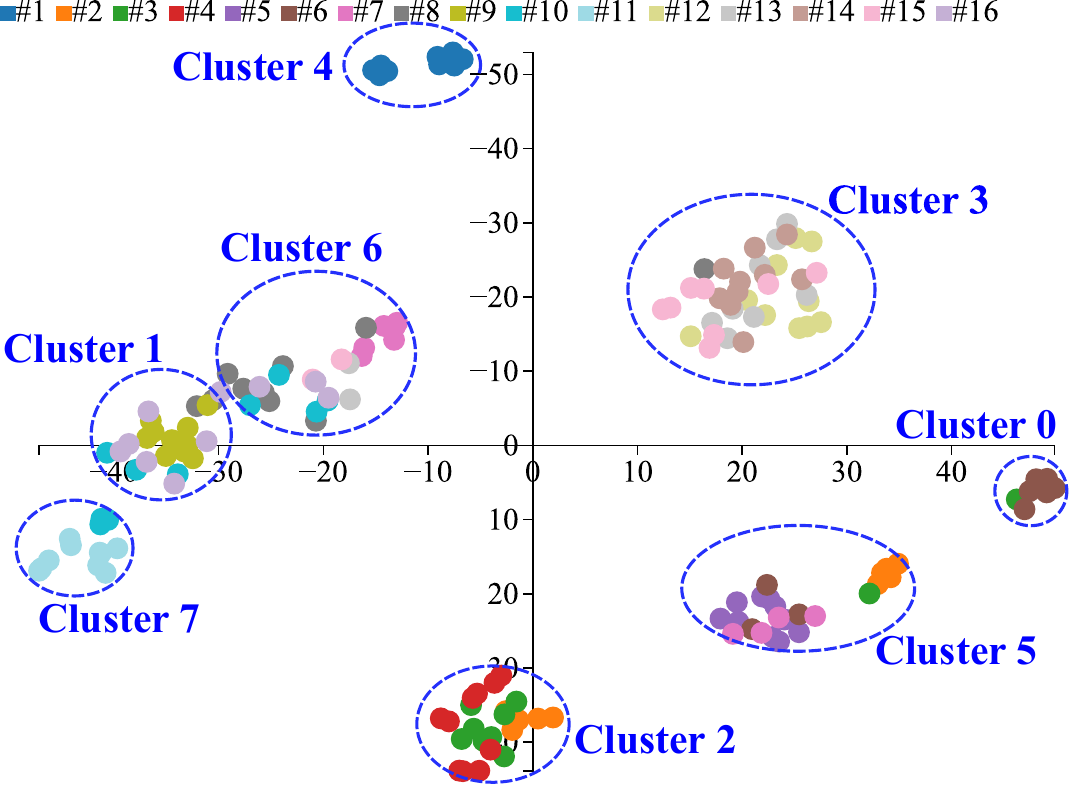}
        \caption{Sensor data representation in 2D plane with TSNE algorithm, when human annotates the IoT states based on their understanding.}
        \label{fig:scatter:collage:1}
    \end{subfigure}
    \caption{Understanding human mental model-based IoT state annotation and sensing model-based IoT state representation with correlation matrix and sensor data visualization in the exploring stage.}
    \label{fig:collage:1}
\end{figure*}

\subsection{Exploring}
\label{sec:iot:probeing:exploring}
As the user interacts with an IoT device to explore its possible internal states, the side-channel information of this IoT device (e.g., power consumption, network traffic, and emanations) is automatically generated and collected. Simultaneously, the user-device interactions are recorded, which may cause the IoT device to transit from one state to another. Given each IoT device has a finite set of states and possible transitions between states, a Finite State Machine (FSM) can be used to effectively represent the IoT device's states and the transition events between states. Specifically, each node in the FSM represents the IoT state and the edge connecting two nodes represents the transition event. This finite state machine can not only show the IoT device's states and their transitions but also clearly help users monitor the IoT device's states over time.

Aiming at probing the internal states of the IoT device, we prompt the user to annotate each state following every interaction with the IoT device. While a wide range of states are explored, the annotations of these states depend on the user's understanding based on their observations of the IoT device's responses and their interactions. For instance, while interacting with a Google Home smart speaker, one user might annotate the states as 'question-answering' and 'music-playing' when asking about the weather and playing a song, respectively. Another user, however, might annotate both states simply as 'responding'.



 Therefore, even though more states are probed, they suffer from being less resilient to noise and misannotation, which can introduce confusion on the IoT device's internal state annotation and verification in the later stage. As a result, the finite state machine is over-semantic or less-semantic. Fig.~\ref{fig:cm:collage:1} shows the correlation matrix derived from the sensing model and the human mental model. Each element in the correlation matrix indicates the percentage of the data points of human-annotated states that belong to the sensing model-generated cluster. As we can see, the human mental model has generated multiple IoT states that are more than the number of IoT states indicated by the sensing model. In other words, each cluster contains multiple human-annotated states. So, there is a misinterpretation of the IoT device's internal states for the sensing model and human mental model.  To be more visual, Fig.~\ref{fig:scatter:collage:1} shows the processed sensor data in a 2D plane with the TSNE algorithm~\cite{wattenberg2016use}, indicating the IoT states annotated by the human model are over-semantic, as each sensing model-generated cluster may contain multiple human-annotated states with the same semantics.

\begin{figure*}[t!]
    \centering
    \begin{subfigure}[t]{0.5\textwidth}
        \centering
        \captionsetup{width=0.96\textwidth}
        \includegraphics[width=\textwidth]{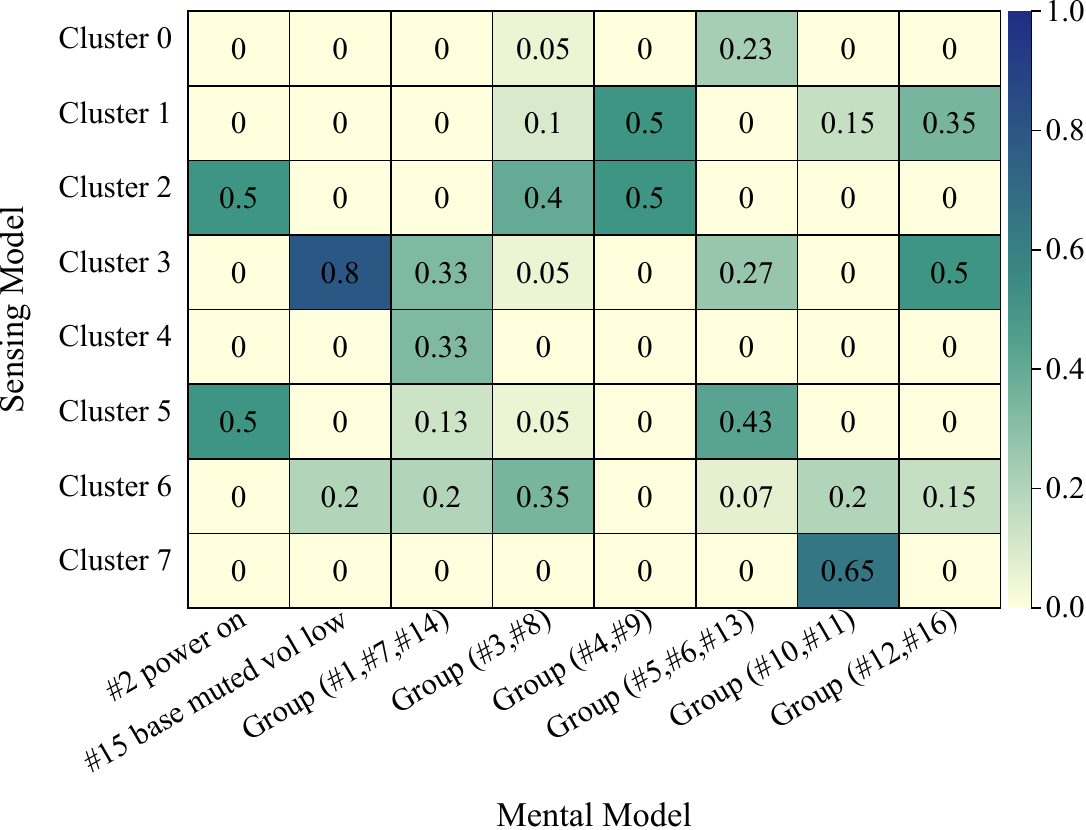}
        \caption{Correlation matrix of the IoT states derived from the sensing model and human mental model using transition event as a hint for state annotation, where each element indicates the percentage of the human-annotated states on the cluster generated by the sensing model.}
        \label{fig:cm:collage:2}
    \end{subfigure}%
    ~ 
    \begin{subfigure}[t]{0.5\textwidth}
        \centering
        \captionsetup{width=0.96\textwidth}
        \includegraphics[width=\textwidth]{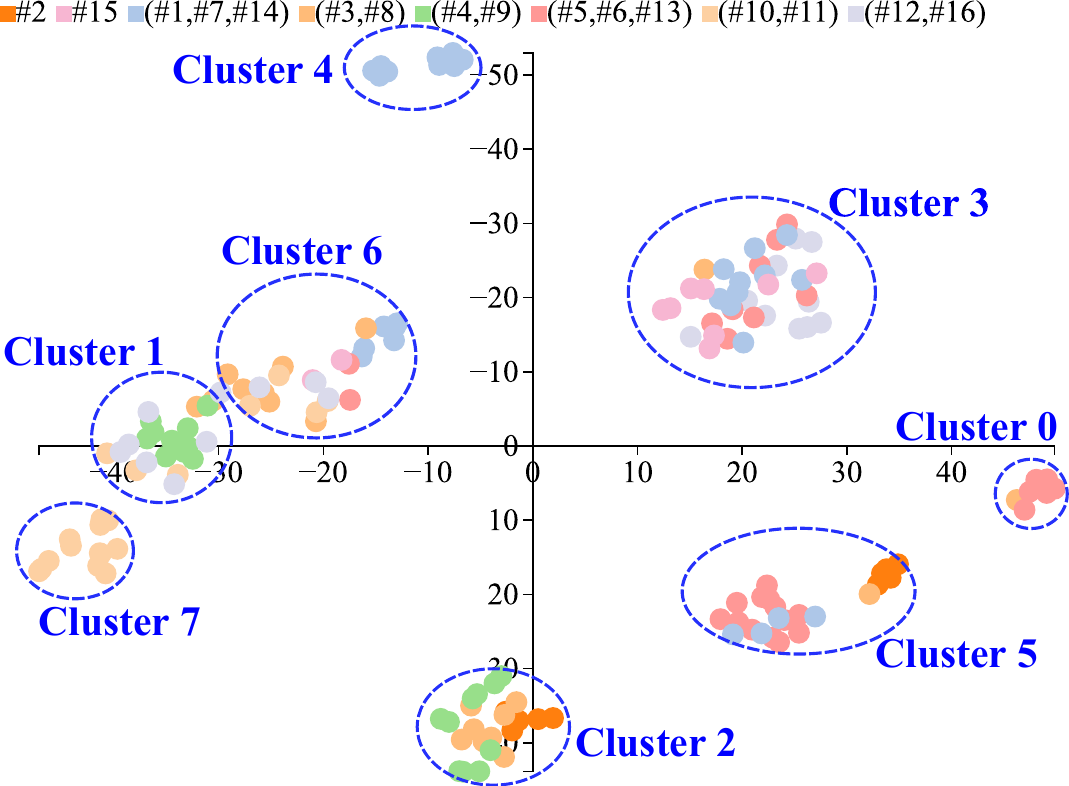}
        \caption{Sensor data representation in the 2D plane with TSNE algorithm, when the states are merged based on the transition events.}
        \label{fig:scatter:collage:2}
    \end{subfigure}
    \caption{The IoT states merging based on the transition states can be helpful for the semantic IoT states annotation.}
    \label{fig:collage:2}
\end{figure*}

\subsection{Modeling}
\label{sec:iot:probeing:modeling}
Before the user engages in refining the FSM, we utilize two types of information to provide a better starting point for the user. First, we leverage the transition events between states. Given that the same transition events are likely to lead to the same state, they can be used as a hint to identify the IoT state with the same semantics. For instance, whenever an user says a keyword (e.g., 'OK Google') to start interacting with a Google Home smart speaker, it may always enter the same state, waiting for further commands. Therefore, we merge the states derived from the same transition event to reduce semantics redundancy. Second, we process the sensing data collected during exploration to develop a sensing model as described in Sec~\ref{sec:iot:sensing}. This sensing model offers insights from the statistical perspective and classifies initial states into different clusters.
  

However, while the sensing model itself could serve as a direct characterization of the IoT device’s internal states, it is challenging for an user to understand the meaning of internal state clusters without annotation, which further hinders probing processes. Moreover, the FSM still suffers from inaccurate state representation. This is because the same transition events may lead to different states, meaning the FSM may not accurately correspond to the actual internal states of the IoT device. Fig.~\ref{fig:cm:collage:2} shows the correlation matrix of the IoT device's states derived from the sensing model and human mental model after merging the states derived from the same transition event. As we can see, despite the number of human-annotated states decreasing due to the merging of states, merged states are still scattered across different clusters, which can also be demonstrated through the scatter plot of the sensor data with the TSNE algorithm in Fig.~\ref{fig:scatter:collage:2}.

\begin{figure*}[t!]
    \centering
    \begin{subfigure}[t]{0.5\textwidth}
        \centering
        \captionsetup{width=0.96\textwidth}
        \includegraphics[width=\textwidth]{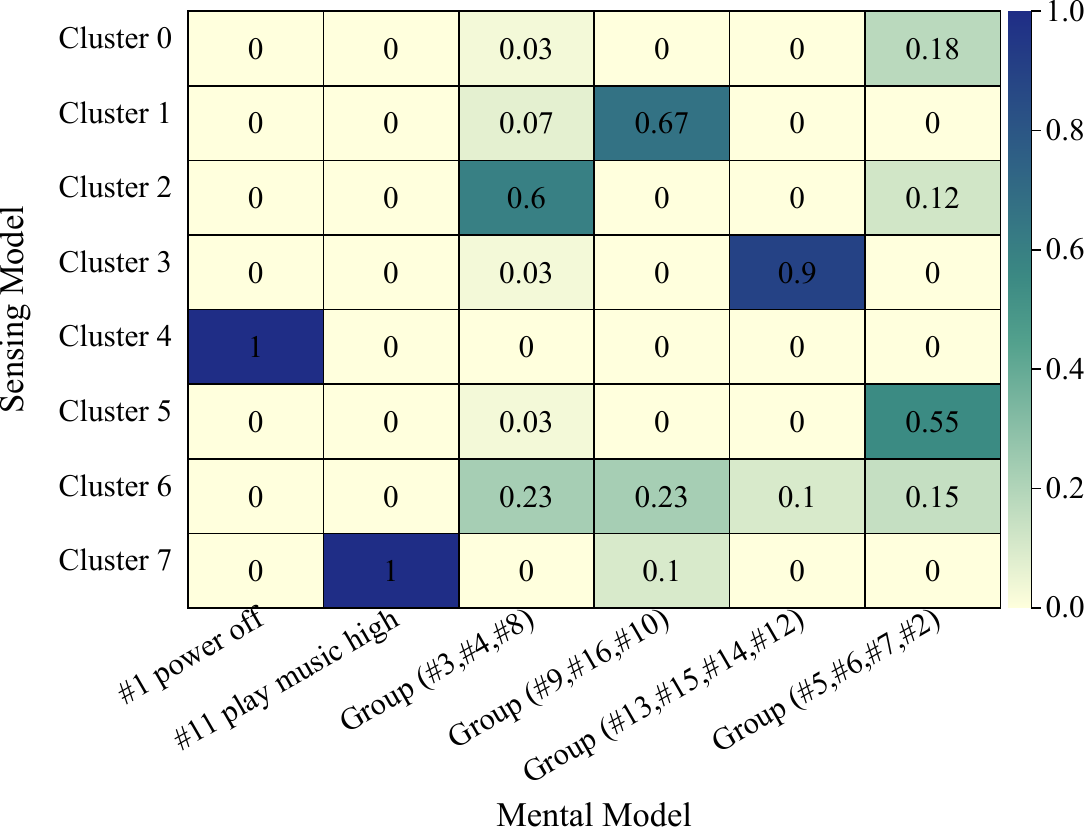}
        \caption{Correlation matrix of the IoT states derived from the sensing model and human mental model after making all collages, where each element indicates the percentage of the human-annotated states in the cluster generated by the sensing model.}
        \label{fig:cm:collage:3}
    \end{subfigure}%
    ~ 
    \begin{subfigure}[t]{0.5\textwidth}
        \centering
        \captionsetup{width=0.96\textwidth}
        \includegraphics[width=\textwidth]{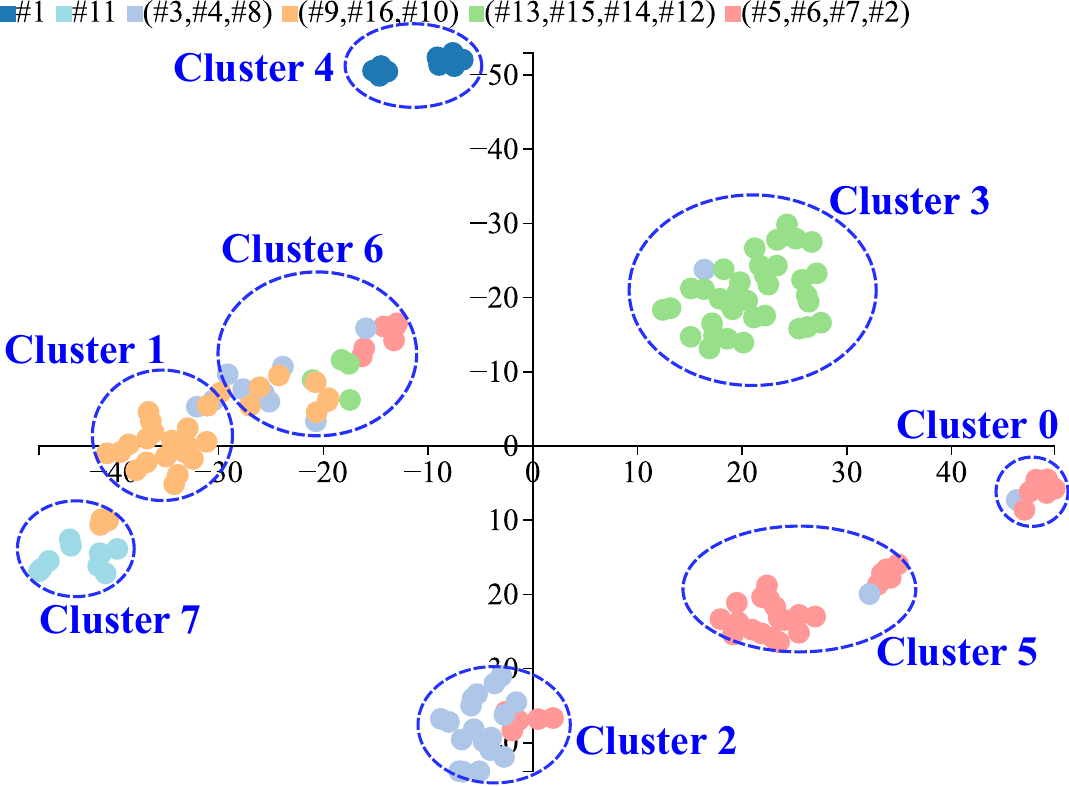}
        \caption{Sensor data representation in the 2D plane with TSNE algorithm after human finishes making all the collages.}
        \label{fig:scatter:collage:3}
    \end{subfigure}
    \caption{By fusing the sensing model with the mental model for collaging, the states of generated FSM show high coherence. This indicates that human understanding aligns well with the sensing data derived from the sensing model.}
    \label{fig:collage:3}
\end{figure*}

\subsection{Collaging}
\label{sec:iot:auditinging:collaging}

To further accurately and semantically probe the internal states of the IoT device, we leverage the sensing model since it can indicate the IoT states through the sensor data representation. As such, the sensor data representing the IoT states with the same semantics can formulate a cluster and further be collaged as one state.

To this end, we integrate the sensor data representation from the sensing model with the annotations and transition events from the mental model into a visual and interactive display. This display illustrates the relationship between mental model-introduced states and sensing model-introduced states through a correlation matrix and a sensor data representation. Along with this display, we also provide context information in our user interface (see Fig.~\ref{fig:ui}), which includes the recording of state information and transition events, to help users with state annotation. By interacting with the user interface, the user can leverage the correlation matrix and sensor data representation from both sensing and mental models to further make a collage of the annotated states. Thus, the collages of states remain semantically understandable to users while becoming more reliable with the aid of underlying low-level data information.

Fig.~\ref{fig:cm:collage:3} shows the correlation matrix of the IoT states derived from the sensing model and human mental model after the user makes a collage. As we can see, the number of states derived from the sensing model is close to the number of states annotated by humans. This is because each sensing model-derived state has a corresponding human mental model-derived state. This can also be demonstrated through Fig.~\ref{fig:scatter:collage:3} showing the scatter plot of the sensor data with the TSNE algorithm after making a collage. As we can see, the clusters derived from the sensing model mainly consist of one state indicated by the human mental model. As a result, the finalized FSM based on the sensing model and human mental model should have semantic states that are capable of representing the IoT device's internal states.

\begin{figure*}
  \centering
  \includegraphics[width=0.8\linewidth]{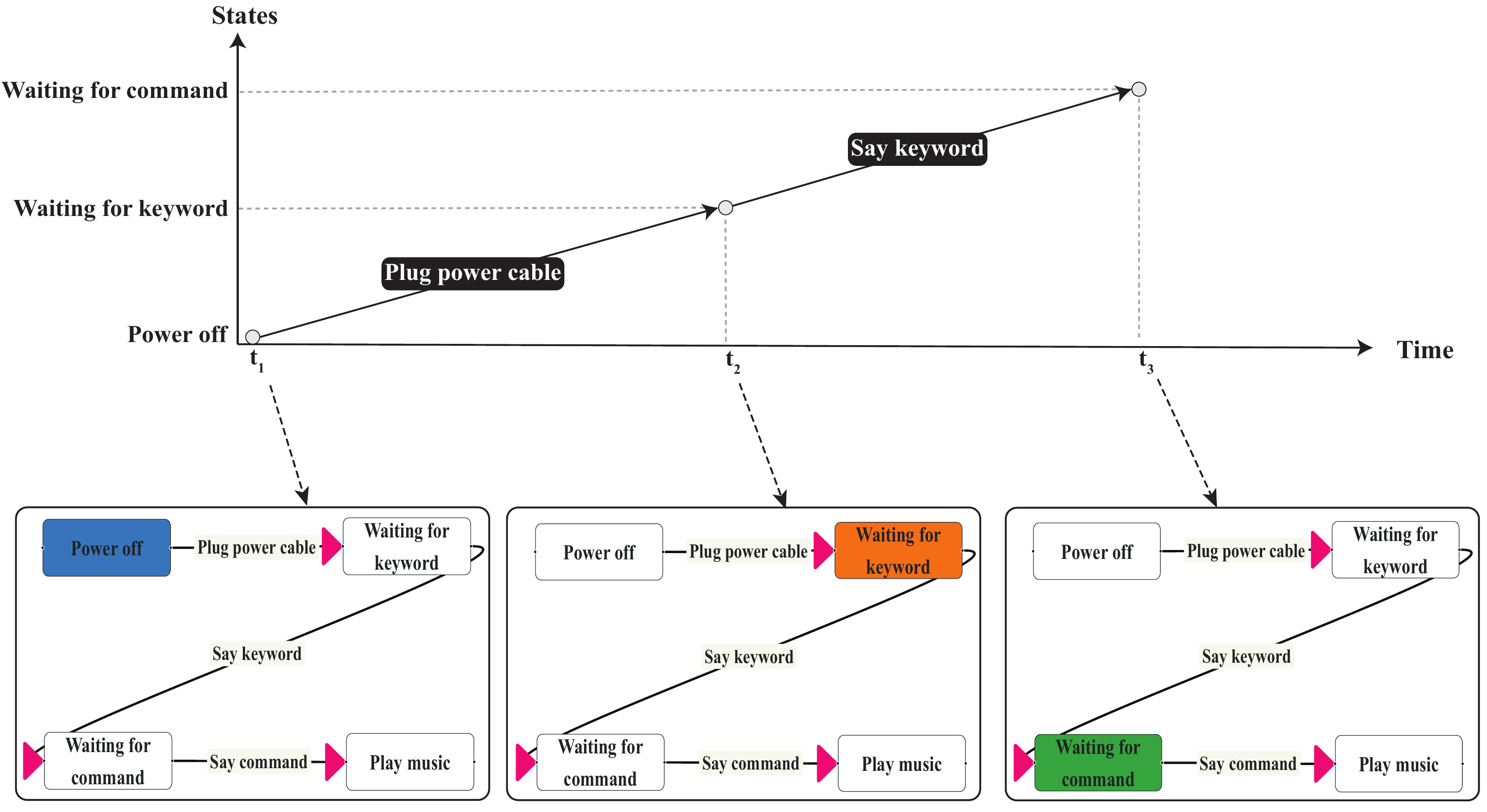}
    \caption{Step-wise verification for the Google Home smart speaker. The top figure showcases the Google Home smart speaker's three states when we interact with it over time. The bottom figure shows the finite state machine of the Google Home smart speaker over time. As we interact with the Google Home smart speaker, its state changes over time which is indicated by the colored node in the finite state machine.}
    \label{fig:aiy:verification}
\end{figure*}

\subsection{Verifying}
\label{sec:iot:auditing:verifying}

After the user extensively exploit the IoT device's internal states, \sysname can generate a finite state machine of this IoT device. As this IoT device is deployed in the physical environment, its internal states should be indicated through the generated finite state machine. To demonstrate the efficiency of the generated finite state machine, we need to verify that \sysname can accurately probe the IoT device's state with the well-fused sensing model and human model obtained in the above sections.

To do so, \sysname utilizes the generated finite state machine to train a classifier. Specifically, the statistical features derived from Sec~\ref{sec:iot:probeing:modeling} are used as data inputs and the annotations of the states serve as labels. During the verification, as the user interacts with the IoT device over time, the generated side-channel information can be used to predict the current IoT state with the well-trained classifier. Fig.~\ref{fig:aiy:verification} showcases the step-wise verification when the user interacts with the Google Home smart speaker.  The top figure shows the state transition of the Google Home smart speaker during the interaction and the bottom figure shows the over-time variation of the Google Home smart speaker's finite-state machine. As we can see, when the speaker is powered up, it enters the waiting-for-keyword state. After we say the keyword (i.e., 'OK Google'), it enters the waiting-for-command state. After we ask it to play music (i.e., say a command), it enters the playing-music state. During this process, the colored node in the finite state machine generated by the sensing model and human mental model indicates the Google Home smart speaker's current state and the other states are indicated by the white nodes.

%% file: ubicomp_parts/p5-implementation.tex
\section{Implementation}
\label{sec:impl:eval}

\vspace{0.1cm}\noindent\textbf{Hardware and Software.} 
\sysname utilizes a combination of power consumption, network traffic, and emanations to sense the internal states of IoT devices. The experimental setup is shown in Fig.~\ref{fig:teaser}.
We use a signal hound~\cite{signal_hound} for spectrum sensing to extract the emanations from the device. We use a power sensor (i.e., ACS172~\cite{acs_172}) connected with the Arduino to measure the real-time power consumption of the IoT device (e.g., Google Home smart speaker~\cite{google_home}, Google AIY voice kit~\cite{google_aiy_voice_kit}). 
We use TShark~\cite{combs2012tshark} to collect real-time network traffic data from the IoT device. The data is streamed to a desktop, where features are extracted in real-time and used to run the sensing model and human mental model for the IoT device's internal state probing. To have a well-trained machine-learning model for IoT state probing, we collect 100 measurements across three side-channel information for each IoT device's state. Then, we simply split the collected dataset into 80$\%$ for training and $20\%$ for testing. We evaluate the performance of our system with two white-box IoT devices (i.e., Google AIY voice kit and vision kit) and one black-box IoT device (i.e., Google Home smart speaker). In our evaluation, we explored three states for the Google AIY vision kit and five states for the Google AIY voice kit. Since we conducted a user study to explore the states of the Google Home smart speaker with 10 volunteers, the explored IoT states should be different for different volunteers depending on their understanding of the IoT device's states. 

\begin{figure*}
  \centering
  \includegraphics[width=\linewidth]{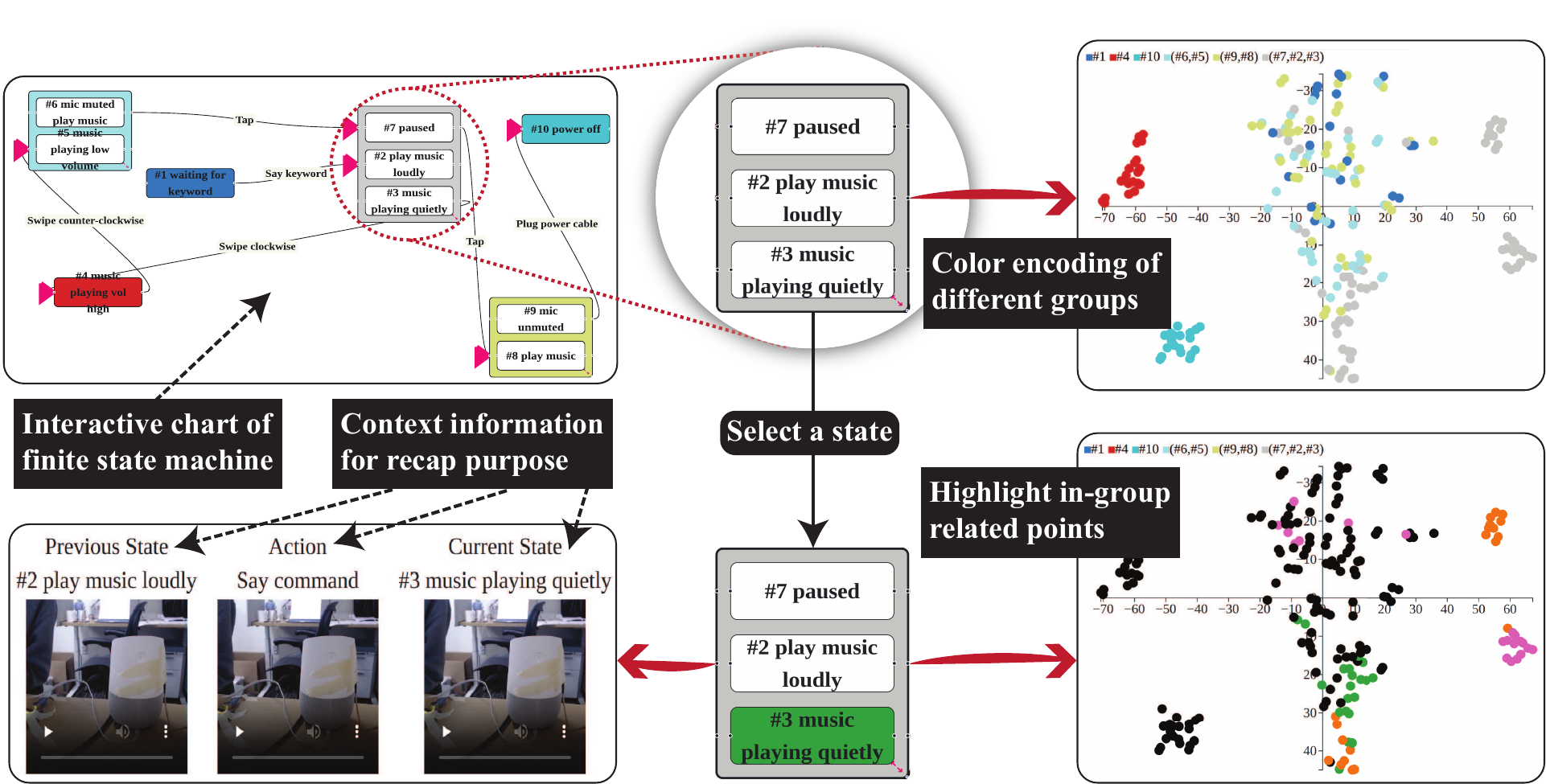}
    \caption{\sysname's graphical user interface consists of an interactive chart workspace for users to annotate and collage the states, context information, and visually plotted sensor data to assist user's state annotation and collaging. The zoomed-out figure shows the state collaged by humans based on the sensor data representation and context information.}
    \label{fig:ui}
\end{figure*}

\vspace{0.1cm}\noindent\textbf{User Interface Design.}
To assist the users in probing the internal states of IoT devices, as shown in Fig.~\ref{fig:teaser}, we develop a graphical user interface (GUI) which is illustrated in Fig.~\ref{fig:ui} consisting of an interactive chart workspace, context information, and visually plotted sensor data to assist user's state annotation and collaging. We utilize the React framework~\cite{react} in JavaScript for the front-end design and FastAPI~\cite{fastapi} in Python for the back-end design. For the front-end design, we employ D3.js~\cite{d3js} and React Flow~\cite{reactworkflow} to illustrate the finite state machine, correlation matrix, and scatter plot that can showcase the data distribution in the 2D plane to assist the user's state annotation and probing.

\vspace{0.1cm}\noindent\textbf{Experimental Settings for White-box IoT State Sensing.} We evaluate the performance of the white-box IoT state sensing with Google AIY vision and voice kits. Since we can program these two kits, we can obtain the ground-truth IoT states based on what pieces of the codes are executed. For the sake of simplicity, we obtained the five ground-truth states of the Google AIY voice kit and three ground-truth states of the Google AIY vision kit. For each IoT state of Google AIY vision and voice kit, we collect the side-channel information 100 times for training and testing. We evaluate the performance of the IoT states probing with precision, recall, F1 score, and confusion matrix using DBSCAN, GMM, and k-means.

\vspace{0.1cm}\noindent\textbf{Experimental Settings for Black-box IoT State Sensing.} We evaluate the performance of the black-box IoT state probing with commercial off-the-shelf Google Home smart speaker through a user study. More details can be found in Sec.~\ref{subsec:user:study}.

%% file: ubicomp_parts/p6-results.tex
\section{Experimental Results}
\label{sec:results}

\begin{figure}
  \centering
\includegraphics[width=0.3\linewidth]{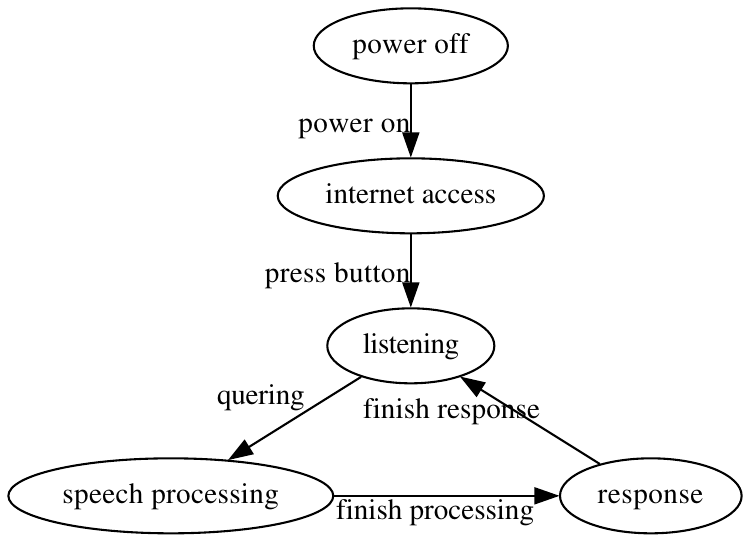}
    \caption{Finite state machine for Google AIY voice kit during the interaction.}
    \label{fig:aiy:fsm}
\end{figure}

\begin{figure*}
\centering
\begin{minipage}{0.33\textwidth}
\captionsetup{width=0.96\textwidth}
\centering
    \includegraphics[width=\linewidth]{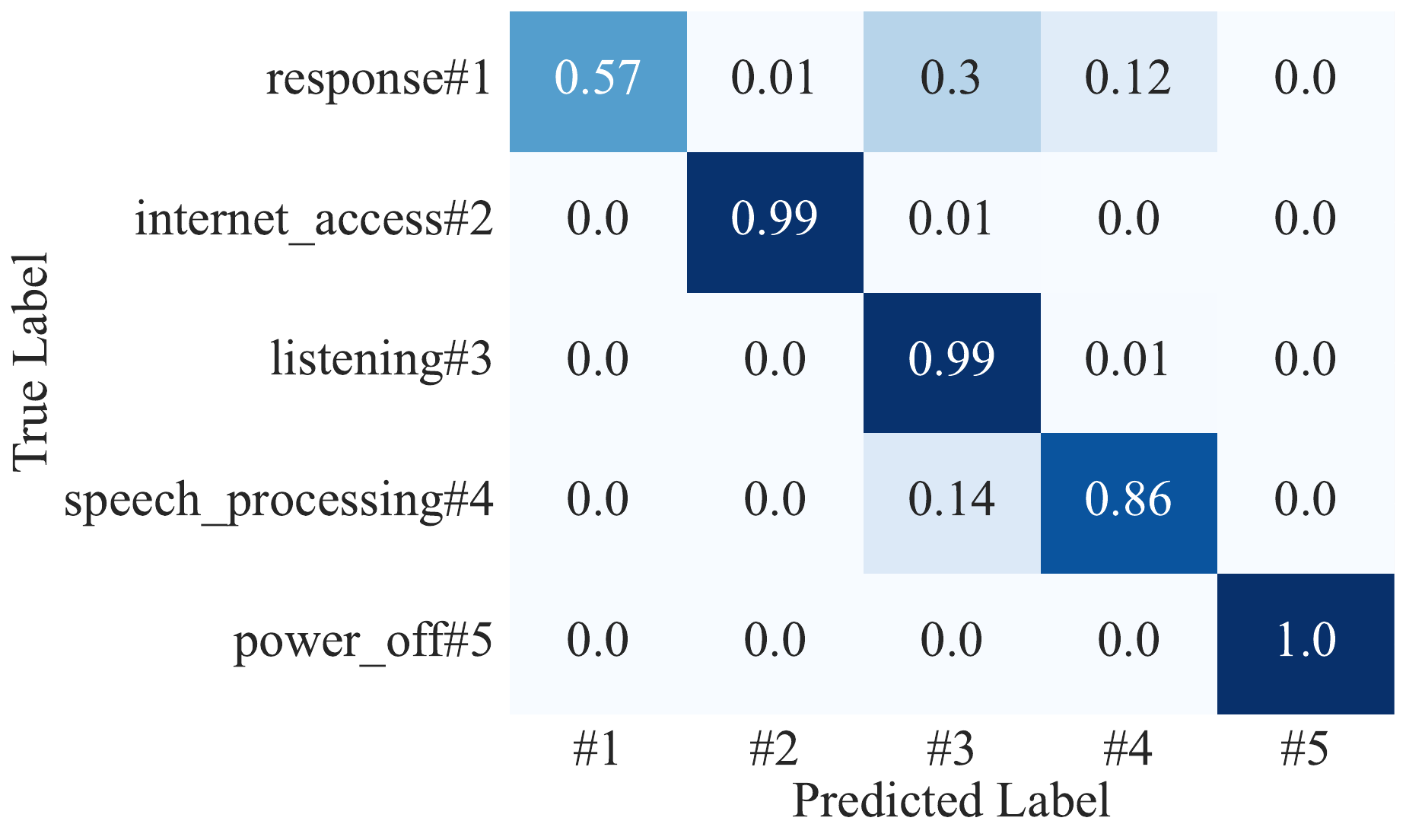}
 \caption{Multimodal sensor fusion-based IoT state detection with k-means for Google AIY voice kit.}
    \label{fig:aiy:multimodal:kmeans}
\end{minipage}%
\begin{minipage}{0.33\textwidth}
\captionsetup{width=0.96\textwidth}
\centering
    \includegraphics[width=\linewidth]{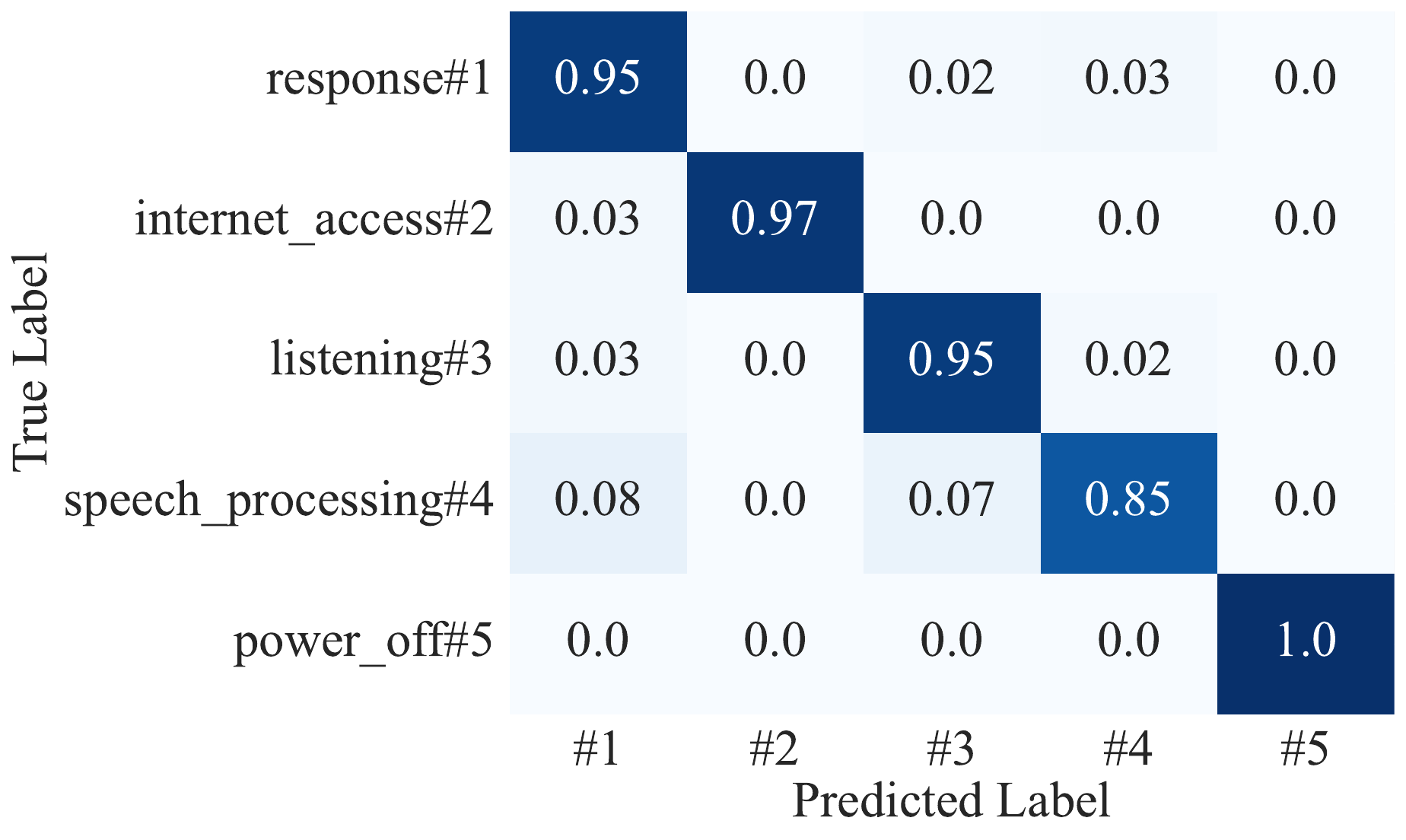}
    \caption{Multimodal sensor fusion-based IoT state detection with DBSCAN for Google AIY voice kit.}
    \label{fig:aiy:multimodal:dbscan}
\end{minipage}%
\begin{minipage}{0.33\textwidth}
\captionsetup{width=0.96\textwidth}
\centering
   \includegraphics[width=\linewidth]{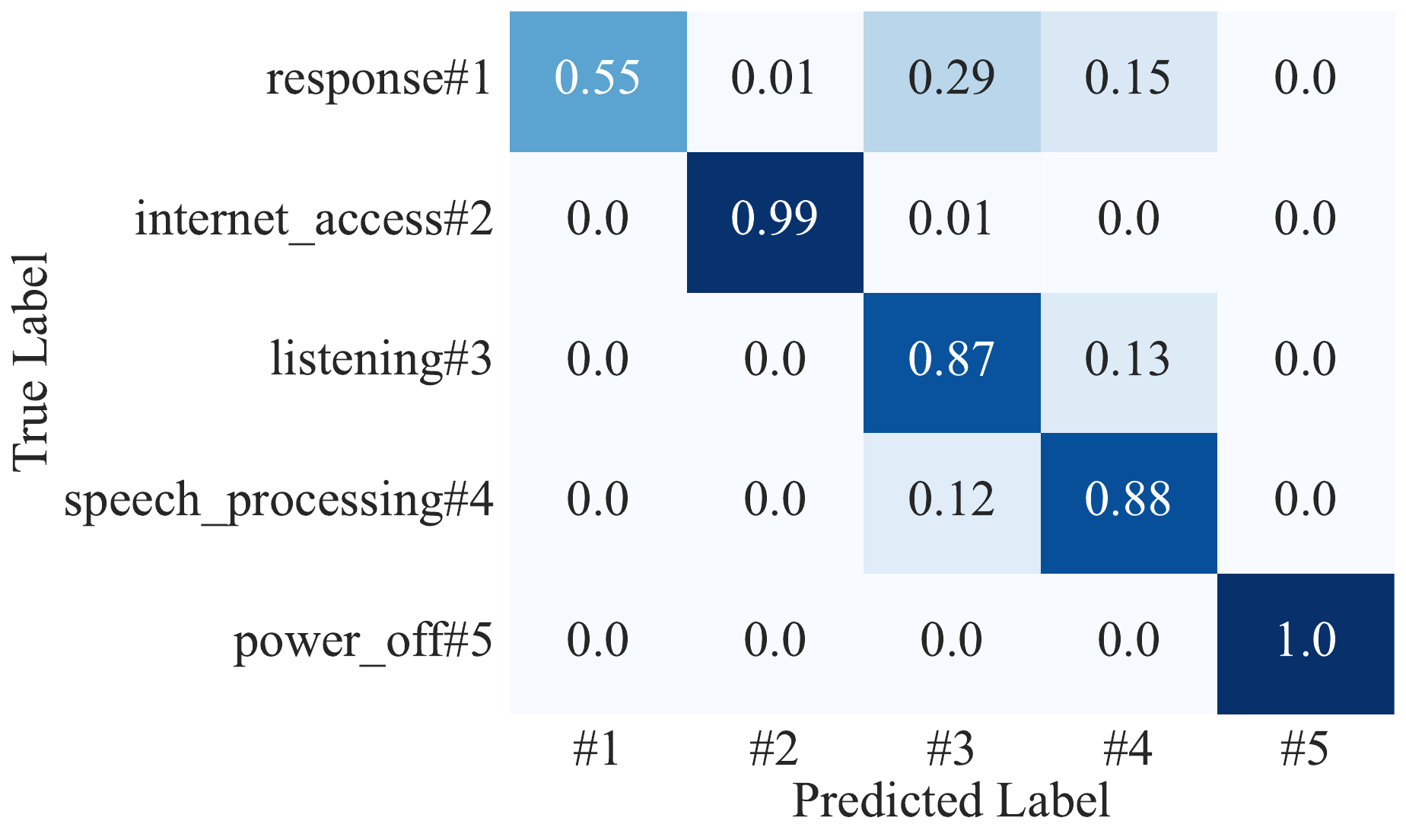} 
    \caption{Multimodal sensor fusion-based IoT state detection with GMM for Google AIY voice kit.}
    \label{fig:aiy:multimodal:gmm}
\end{minipage}
\end{figure*}

\begin{figure*}
\centering
\begin{minipage}{0.5\textwidth}
\captionsetup{width=0.96\textwidth}
\centering
   \includegraphics[width=0.8\linewidth]{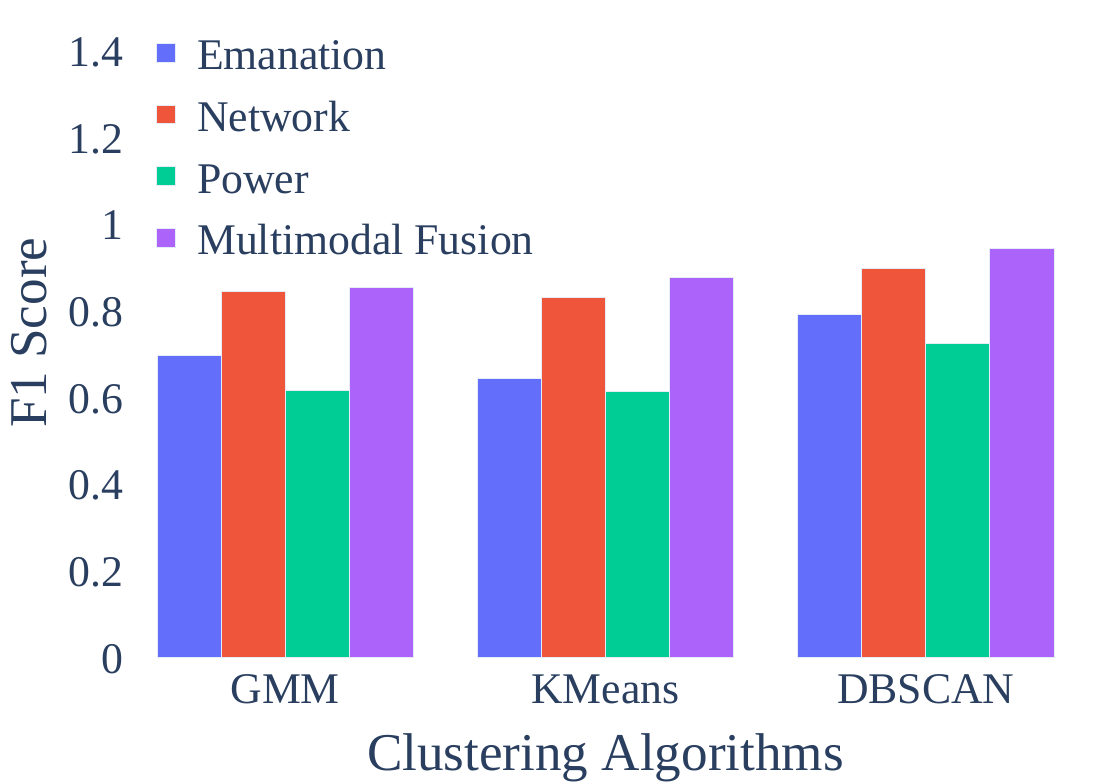}
    \caption{F1 score of IoT state detection using k-means, DBSCAN, and GMM for Google AIY voice kit.}
    \label{fig:aiy:f1score}
\end{minipage}%
\begin{minipage}{0.5\textwidth}
\captionsetup{width=0.96\textwidth}
\centering
   \includegraphics[width=0.68\linewidth]{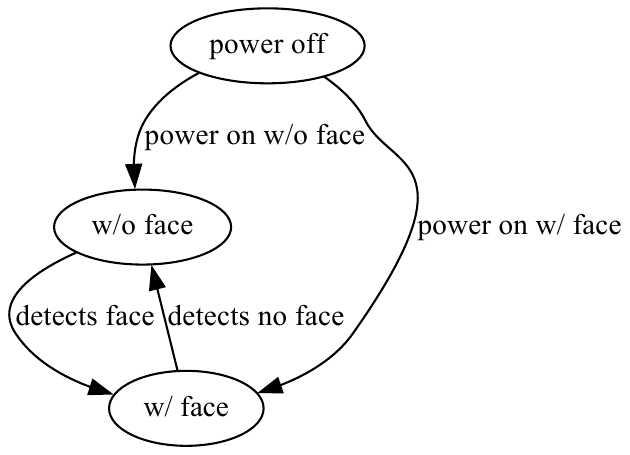}
    \caption{Finite state machine for Google AIY vision kit.}
    \label{fig:aiy:vision:fsm}
\end{minipage}
\end{figure*}

\subsection{White-box IoT Device Probing Evaluation}
To obtain the ground-truth of the internal states of IoT devices, we test \sysname\ with open source hardware (Google AIY voice kit and vision kit). Fig.~\ref{fig:aiy:fsm} illustrates the finite-state machine of the Google AIY voice kit. 
When the Google AIY voice kit is powered up and enters interaction mode, it first accesses the Internet (i.e., Internet access state). Then, it waits for the voice commands/queries (i.e., listening state). After we query the Google AIY voice kit, it enters the speech processing state to understand the commands or queries. Finally, it responds to us by finding the answers from the Internet. 


\begin{figure*}
\centering
\begin{minipage}{0.33\textwidth}
\captionsetup{width=0.96\textwidth}
\centering
    \includegraphics[width=\linewidth]{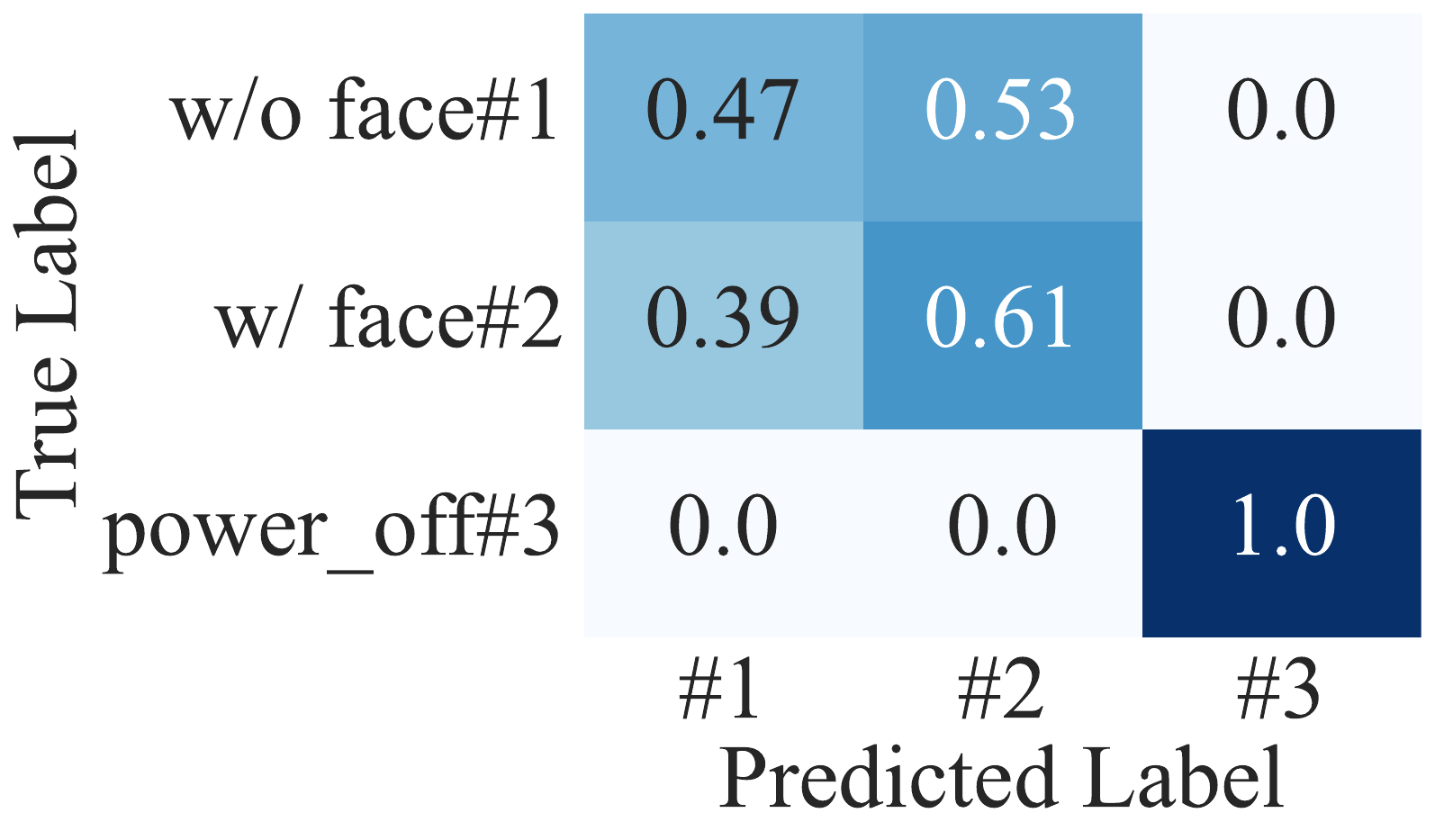}
 \caption{Multimodal sensor fusion-based IoT state detection with k-means for Google AIY vision kit.}
    \label{fig:aiy:vision:multimodal:kmeans}
\end{minipage}%
\begin{minipage}{0.33\textwidth}
\captionsetup{width=0.96\textwidth}
\centering
    \includegraphics[width=\linewidth]{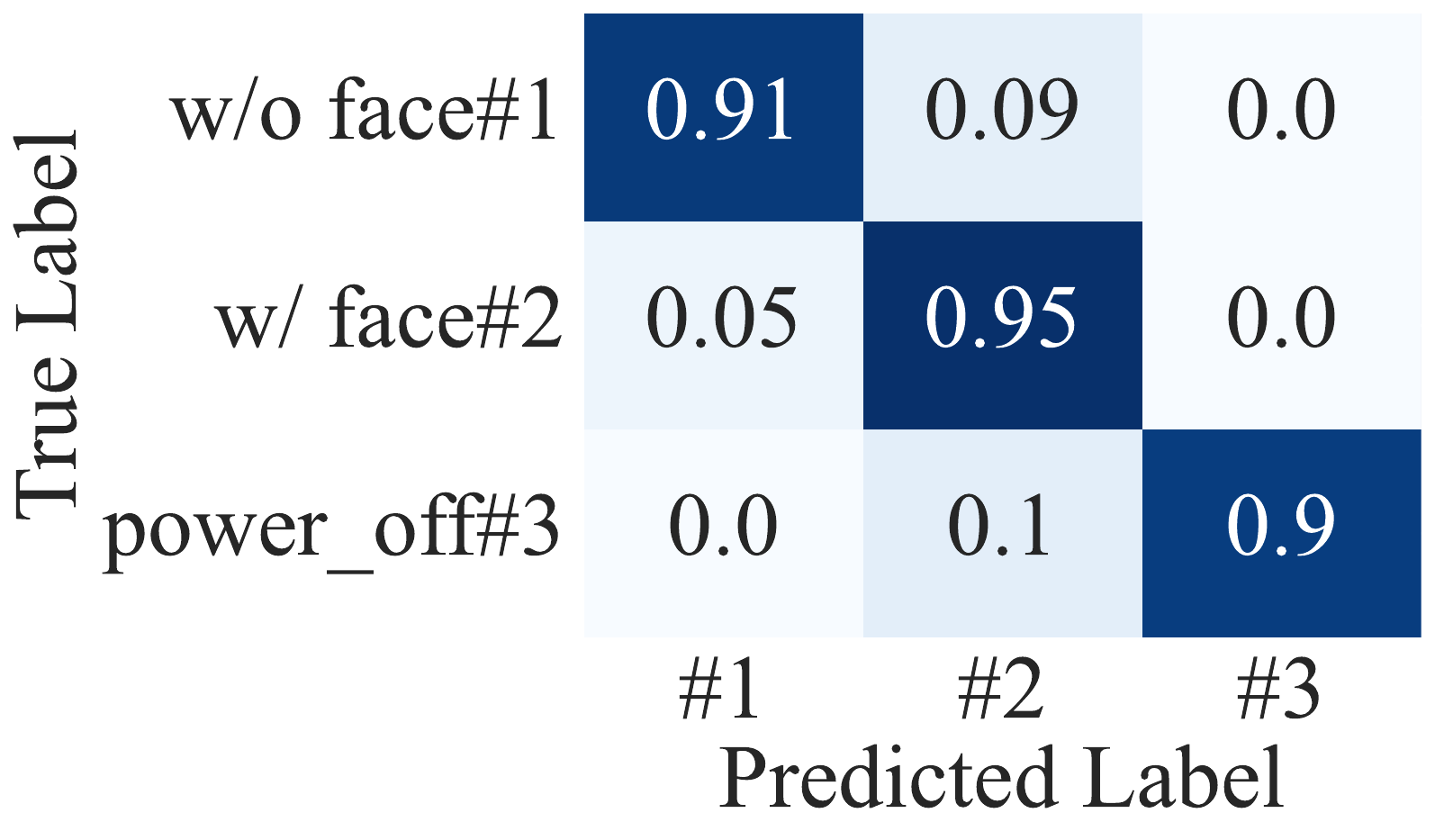}
    \caption{Multimodal sensor fusion-based IoT state detection with DBSCAN for Google AIY vision kit.}
    \label{fig:aiy:vision:multimodal:dbscan}
\end{minipage}%
\begin{minipage}{0.33\textwidth}
\captionsetup{width=0.96\textwidth}
\centering
   \includegraphics[width=\linewidth]{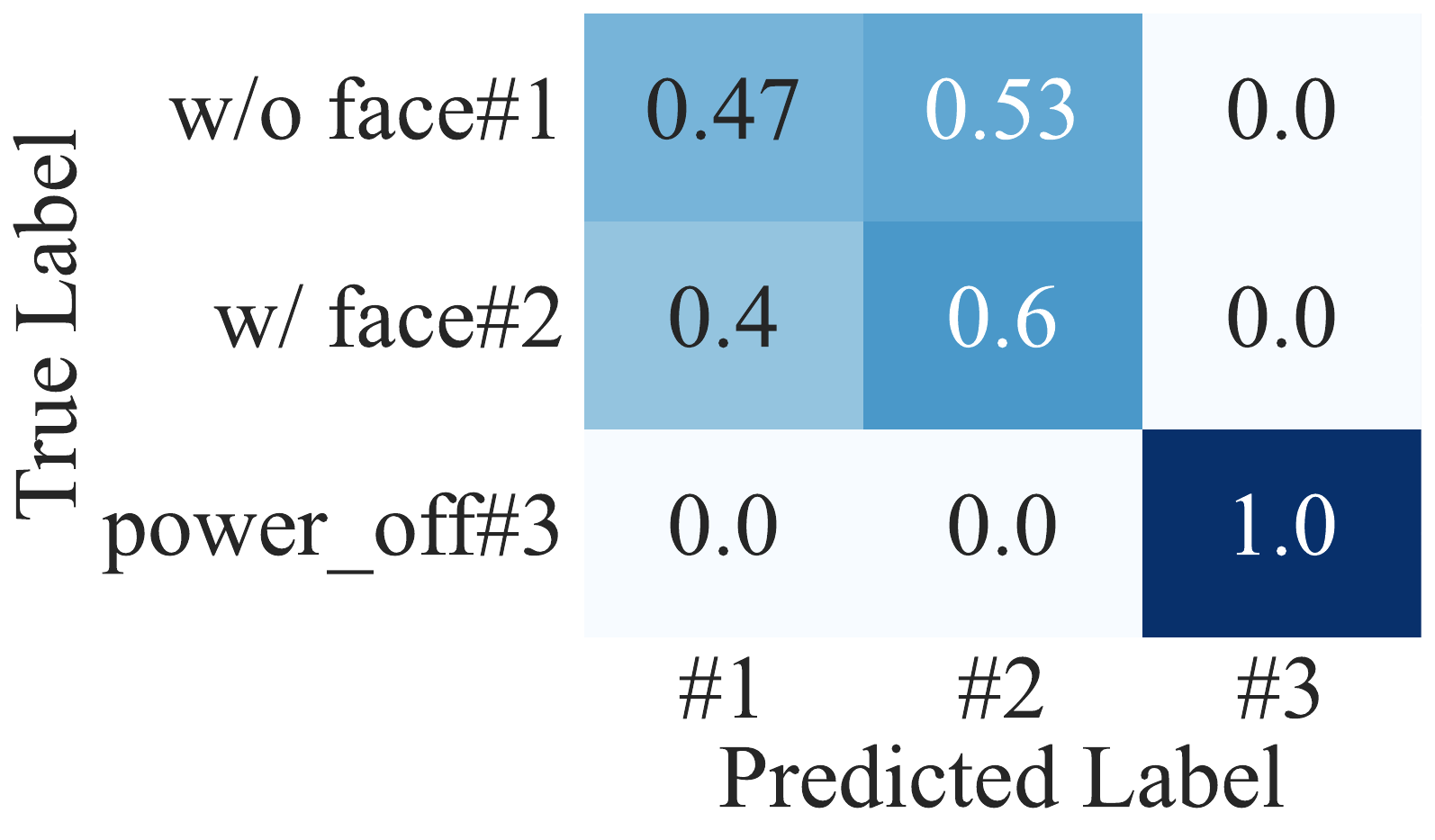} 
    \caption{Multimodal sensor fusion-based IoT state detection with GMM for Google AIY vision kit.}
    \label{fig:aiy:vision:multimodal:gmm}
\end{minipage}
\end{figure*}

\begin{figure*}
\centering
\begin{minipage}{0.35\textwidth}
\captionsetup{width=0.96\textwidth}
\centering
   \includegraphics[width=\linewidth]{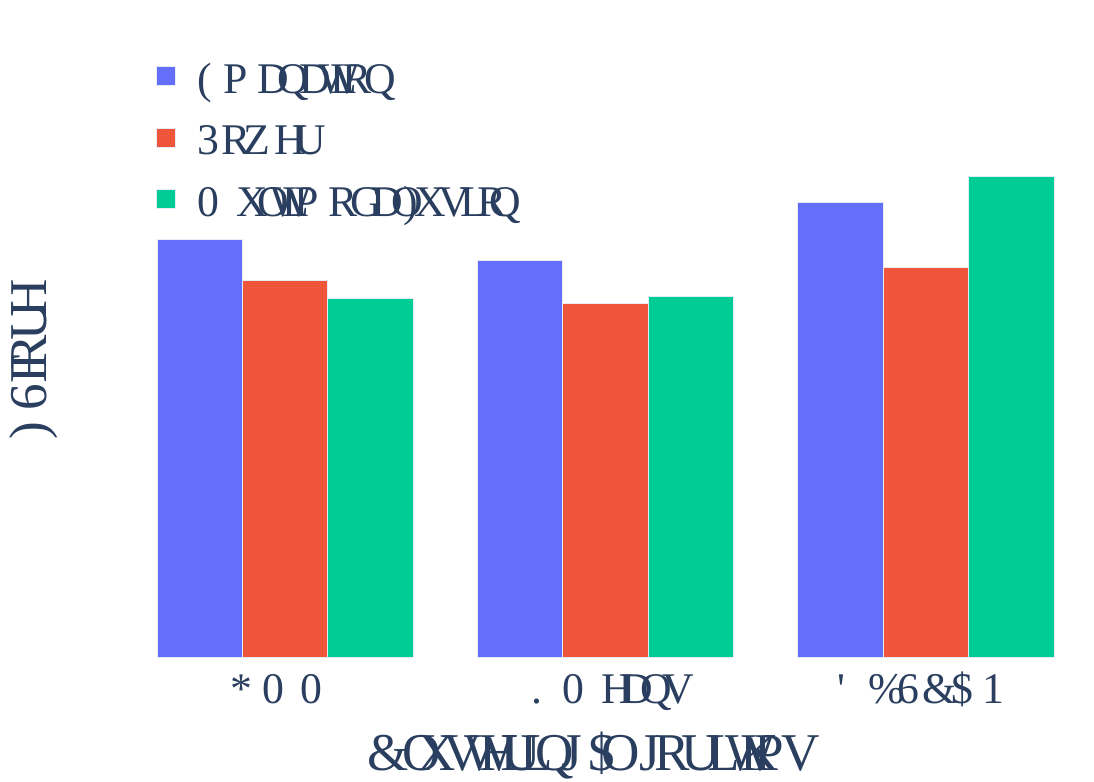}
    \caption{F1 score of IoT state probing using k-means, DBSCAN, and GMM for Google AIY vision kit.}
    \label{fig:aiy:vision:f1score}
\end{minipage}%
\begin{minipage}{0.32\linewidth}%
\centering
\captionsetup{width=0.96\linewidth}
\begin{tabular}{|c|c|c|}
\hline
 & \textbf{Precision}          & \textbf{Recall}             \\ \hline
\textbf{GMM}                   & 0.88 & 0.86             \\ \hline
\textbf{KMeans}                & 0.91 & 0.88             \\ \hline
\textbf{DBSCAN}                & 0.95 & 0.94             \\ \hline
\end{tabular}
\captionof{table}{Precision and recall with k-means, GMM, and DBSCAN for Google AIY Voice Kit's state probing.}
\label{table:voice:precision:recall}
\end{minipage}%
\begin{minipage}{0.32\linewidth}
\centering
\captionsetup{width=0.96\linewidth}
\begin{tabular}{|c|c|c|}
\hline
 & \textbf{Precision}          & \textbf{Recall}             \\ \hline
\textbf{GMM}                   & $0.69$ & $0.69$               \\ \hline
\textbf{KMeans}                & $0.69$ & $0.69$ \\ \hline
\textbf{DBSCAN}                & $0.93$ & $0.92$               \\ \hline
\end{tabular}
\captionof{table}{Precision and recall with k-means, GMM, and DBSCAN for Google AIY Vision Kit state probing.}
\label{table:vision:precision:recall}
\end{minipage}
\end{figure*}

\vspace{0.1cm}\noindent\textbf{Performance with Google AIY voice kit}: Fig.~\ref{fig:aiy:multimodal:kmeans}, Fig.~\ref{fig:aiy:multimodal:dbscan}, and Fig.~\ref{fig:aiy:multimodal:gmm} show the confusion matrix for multimodal sensor fusion-based IoT state probing using k-means, DBSCAN, and GMM algorithms for Google AIY voice kit. As we can see, multimodal sensor fusion with k-means, DBSCAN, and GMM algorithms can achieve a high IoT state probing accuracy of around $0.93$. Fig.~\ref{fig:aiy:f1score} shows the F1 score for IoT state probing using K-means, DBSCAN, and GMM algorithms. From the results, it is clear that multimodal sensor fusion-based IoT state probing performs better than power consumption-based, network traffic-based, and emanation-based IoT state detection. Specifically, the multimodal sensor fusion-based IoT state detection with GMM, k-means, and DBSCAN algorithm can achieve an F1 score of $0.86$,  $0.84$, and $0.94$ respectively, which is larger than the network traffic-based, power consumption-based, and emanation-based IoT state detection. This is because multimodal sensor fusion leverages multiple side-channel information for IoT state probing, thereby differentiating the internal IoT state. Table~\ref{table:voice:precision:recall} presents the precision and recall with GMM, k-means, and DBSCAN for Google AIY Voice Kit state detection. As we can see, the best precision and recall are provided by the DBSCAN algorithm for the kit's state detection, which is $0.95$ and $0.94$ respectively. This is because the DBSCAN algorithm can automatically detect the cluster number and handle the arbitrary-shaped clusters.

\vspace{0.1cm}\noindent\textbf{Performance with Google AIY vision kit:} The Google AIY vision kit has three states: power-off state, without face detection state, and with face detection state. The state machine of the Google AIY vision kit is shown in Fig.~\ref{fig:aiy:vision:fsm}. Note that the Google AIY vision kit does not introduce any network traffic, as all the computation is locally in the vision kit. This further motivates us to use a multimodal sensor fusion-based method for internal IoT state probing. Fig.~\ref{fig:aiy:vision:multimodal:kmeans}, Fig.~\ref{fig:aiy:vision:multimodal:gmm}, and Fig.~\ref{fig:aiy:vision:multimodal:dbscan} present the confusion matrix for multimodal sensor fusion-based IoT state detection with k-means, GMM, and DBSCAN for Google AIY vision kit respectively. As we can see,  the classification accuracy for multimodal sensor fusion-based IoT state detection with k-means, GMM, and DBSCAN is $0.69$, $0.69$, and $0.92$ respectively. This is because the DBSCAN algorithm can automatically identify the cluster number and well represent the arbitrary-shaped clusters. Fig.~\ref{fig:aiy:vision:f1score} presents the F1 score of IoT state detection using k-means, DBSCAN, and GMM for the Google AIY vision kit. As we can see, the best F1 score of $0.92$ is provided by the multimodal sensor fusion method with the DBSCAN algorithm. The multimodal sensor fusion-based methods with k-means and GMM algorithms do not perform better than the single model-based methods due to the performance limitations of k-means and GMM algorithms. Table~\ref{table:vision:precision:recall} presents the precision and recall with GMM, k-means, and DBSCAN for Google AIY Vision Kit state detection. As we can see, the DBSCAN algorithm can provide a precision of $0.93$ and a recall of $0.92$ for the vision kit's state detection, which is better than the GMM and k-means algorithms due to its automatic cluster number estimation and arbitrary-shaped clusters characterization.

\begin{table*}
    \centering
    \begin{tabular}{cllcc}
        \# &  Gender& Education Level& \makecell{Experience of\\working with IoT device} & \makecell{Experience with\\smart speaker}\\ \hline
        P1 &  Male& Undergraduate& No & No\\
        P2 &  Female& Undergraduate& No & No\\
        P3 &  Male& Graduate& Yes & No\\
        P4 &  Male& Graduate& No & No\\
        P5 &  Male& Undergraduate& No & Amazon Alexa\\
        P6 &  Male& Graduate& No & No\\
        P7 &  Male& Graduate& Yes& Xiaomi Mi\\
        P8 &  Male& Graduate& No & No\\
        P9 &  Male& Graduate& Yes & Google Home\\
        P10 &  Male& Undergraduate& No & Google Home\\
    \end{tabular}
    \caption{Information of participants: major, research, or interaction experience with IoT devices and smart speakers.}
    \label{tab:user_demographic}
\end{table*}

\subsection{User Study}
\label{subsec:user:study}
To evaluate the effectiveness of the probing process supported by \sysname, we conducted a user study with 10 college students. In our study, we selected the Google Home smart speaker as the black-box IoT device to examine the following research questions:
\begin{itemize}
    \item[\textbf{RQ1}] What are \sysname's effects on users' performance in probing internal states of IoT devices?
    \item[\textbf{RQ2}] What are \sysname's effects on the users' cognitive load of probing internal states of IoT devices?
    \item[\textbf{RQ3}] How do users think about the probing process using \sysname?
\end{itemize}

\subsubsection{Ethical Considerations}
Our study was approved by our institution's IRB. All participants consented to have their data recorded and reported in a scholarly publication. Collected data were anonymized after collection and stored in a private location accessed only by the authors. 

\subsubsection{Procedure}
The user study consists of four steps: pre-interview, system introduction, probing, and post-interview. 
\begin{itemize}
    \item \textbf{Pre-interview.} After the participants signed the consent forms, we first conducted an interview about the participants' background, demographic information, prior experience of using IoT devices, etc.
    \item \textbf{System introduction.} Then, we set up a Google Home smart speaker with \sysname in a typical indoor office environment. We gave each participant a 15-minute tutorial, during which we systematically introduced them to all the functions of \sysname, provided them with an instruction table listing all feasible interactions with the Google Home smart speaker~\cite{google_home_instructions}, and allowed them to try these interactions to become familiar with the device.
    \item \textbf{Probing.} After the participant was familiar with the user interface and system, the participant was presented with the graphical user interface as shown in Fig.~\ref{fig:ui} and instructed to interact extensively with the IoT device. He/she was encouraged to explore a wide range of interactions, annotate and refine the FSM to match the mental model and sensing model, and subsequently verify the generated FSM.
    \item \textbf{Post-interview.} After the probing was done, the participant was asked to complete a NASA TLX test~\cite{index1990results}. We also interviewed the participants about their feelings and rationale of operations based on the participant observation of the probing process.
\end{itemize}


\subsubsection{Participants}
We recruited 10 college students as participants. Table~\ref{tab:user_demographic} enumerates a breakdown of the participant information. Participants are all majoring in STEM fields (i.e., electrical and computer engineering, computer science and engineering, or data science). One participant self-identified as female, and the remaining students identified as male. Participants P1 and P3 had research experience working with IoT devices. Participants P5 and P8 had used Google Home smart speakers. The other participants either had experience with similar devices (i.e., Amazon Alexa smart speaker, Xiaomi Mi smart speaker) or were not familiar with IoT devices.

\begin{table*}
\begin{minipage}{0.5\textwidth}
\captionsetup{width=0.96\textwidth}
\centering
    \centering  \includegraphics[width=\linewidth]{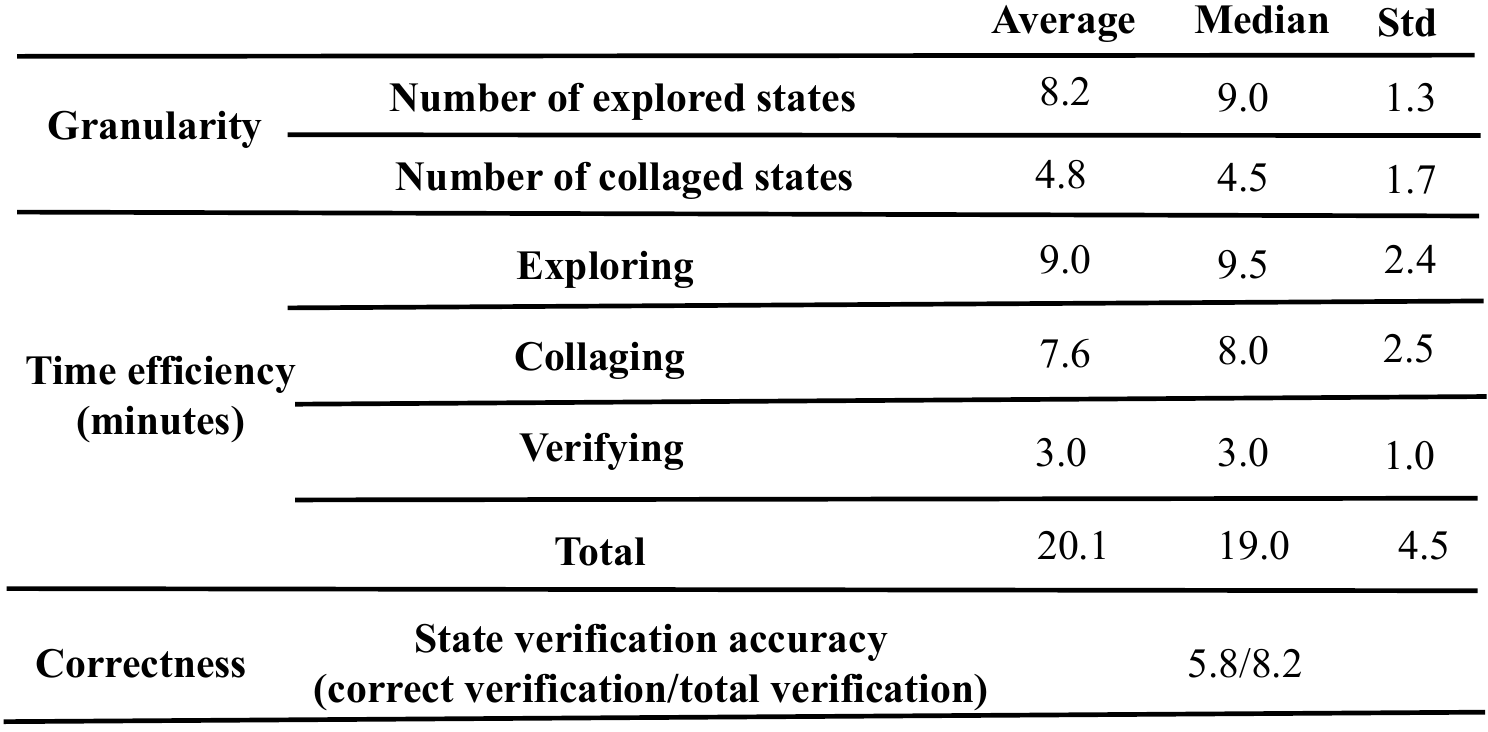} 
 \caption{Evaluation of user study on probing time in different probing stages, number of explored/collaged states, and state probing accuracy.}
\label{tb:quanti_1}
\end{minipage}%
\begin{minipage}{0.5\textwidth}
\captionsetup{width=0.96\textwidth}
\centering
   \includegraphics[width=0.8\linewidth]{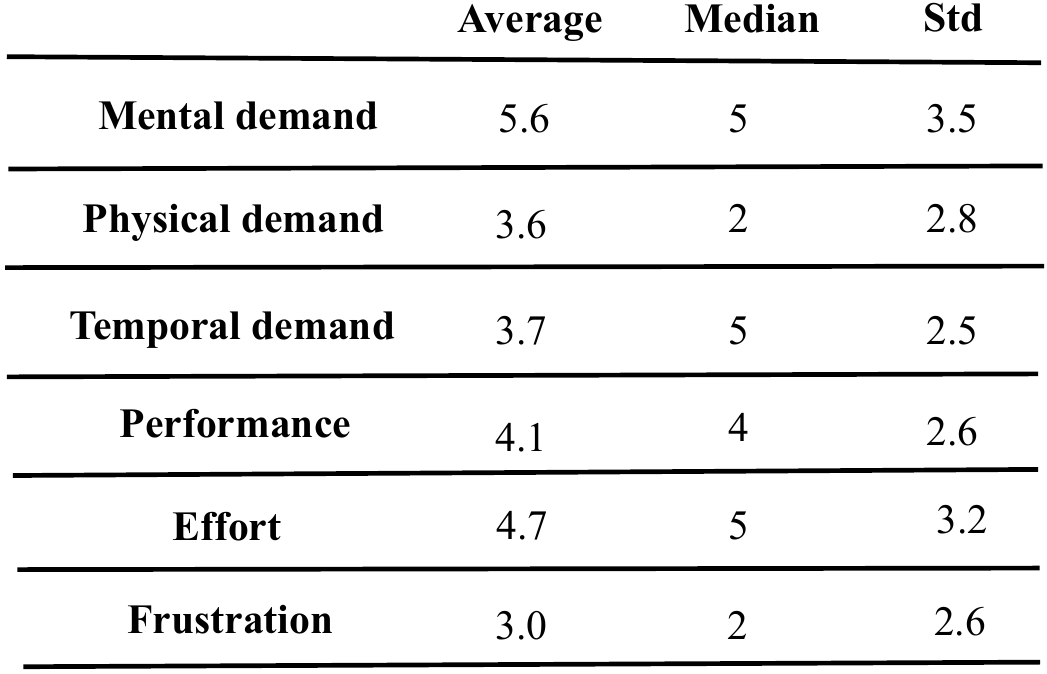}
    \caption{Evaluation of user study with NASA task load index.}
    \label{tb:quanti_2}
\end{minipage}
\end{table*}

\subsubsection{Quantitative Results}
To answer RQ1 and RQ2, we conducted a quantitative analysis.

\vspace{0.1cm}\noindent\textbf{Performance}. Table~\ref{tb:quanti_1} shows the performance of the user study quantitatively. Specifically, we mainly exploit three aspects of the results: (1) quantity of annotated states across different modules, (2) probing time across different modules of \sysname, and (3) accuracy of states probing when the participants use \sysname for IoT states annotation. As we can see from the table, all Participants finished probing the Google Home smart speaker in a relatively short time (i.e., an average time of 20.1s, a median time of 19.0s, a time standard deviation of  4.5s) with internal states probing (i.e., an average time of 4.8s, a median time of 4.5s, a standard deviation time of 1.7s). Furthermore, participants can verify the IoT states with an accuracy of $5.8/8.2$, where 5.8 indicates the average number of correct verified states and 8.2 indicates the average number of states probed by the participants using \sysname.

\begin{figure*}[t!]
    \begin{subfigure}[t]{0.5\textwidth}
        \centering
        \includegraphics[width=\textwidth]{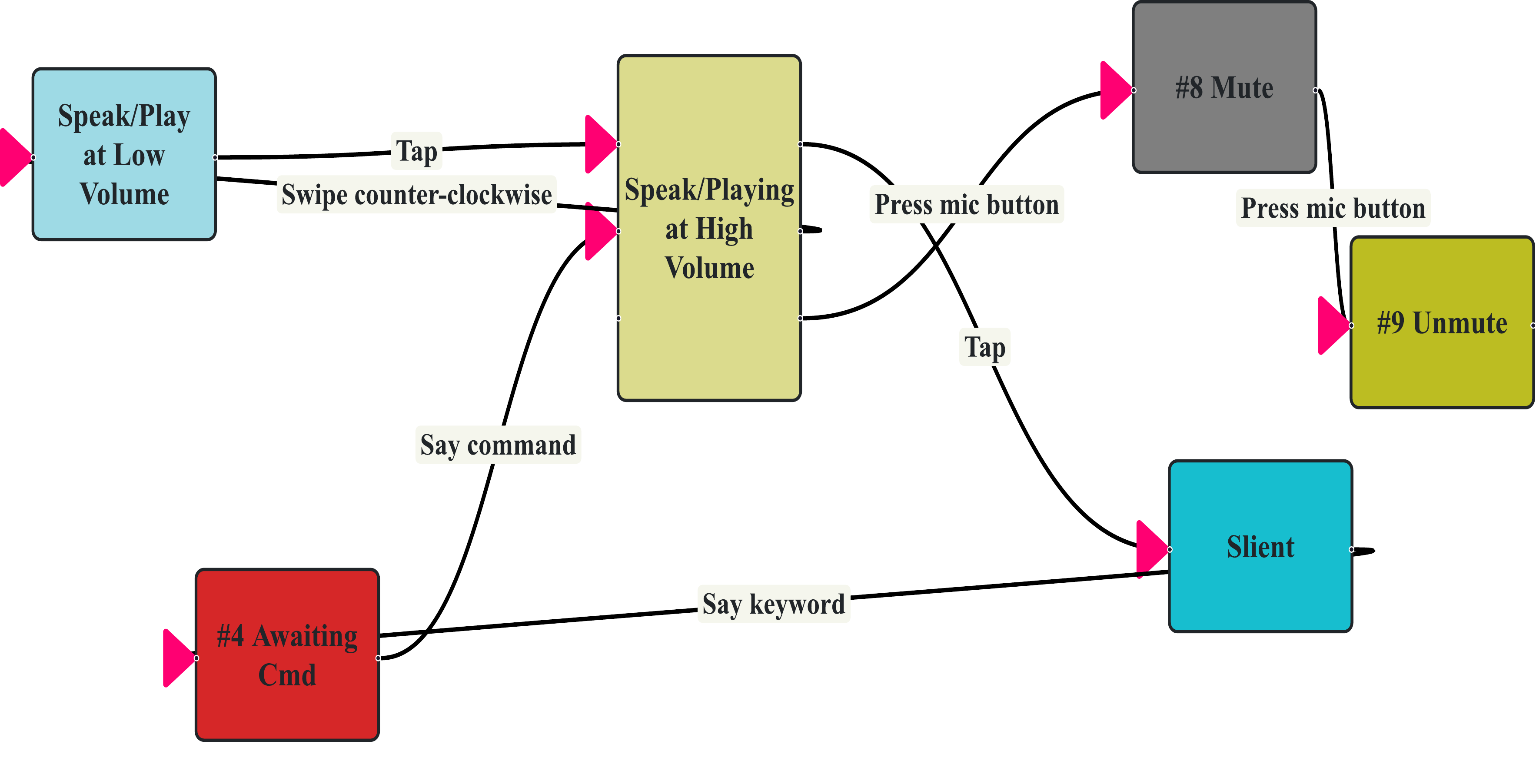}
        \caption{Final FSM from P2.}
        \label{fig:fsm_result:1}
    \end{subfigure}%
    ~
    \begin{subfigure}[t]{0.5\textwidth}
    \centering
    \includegraphics[width=\textwidth]{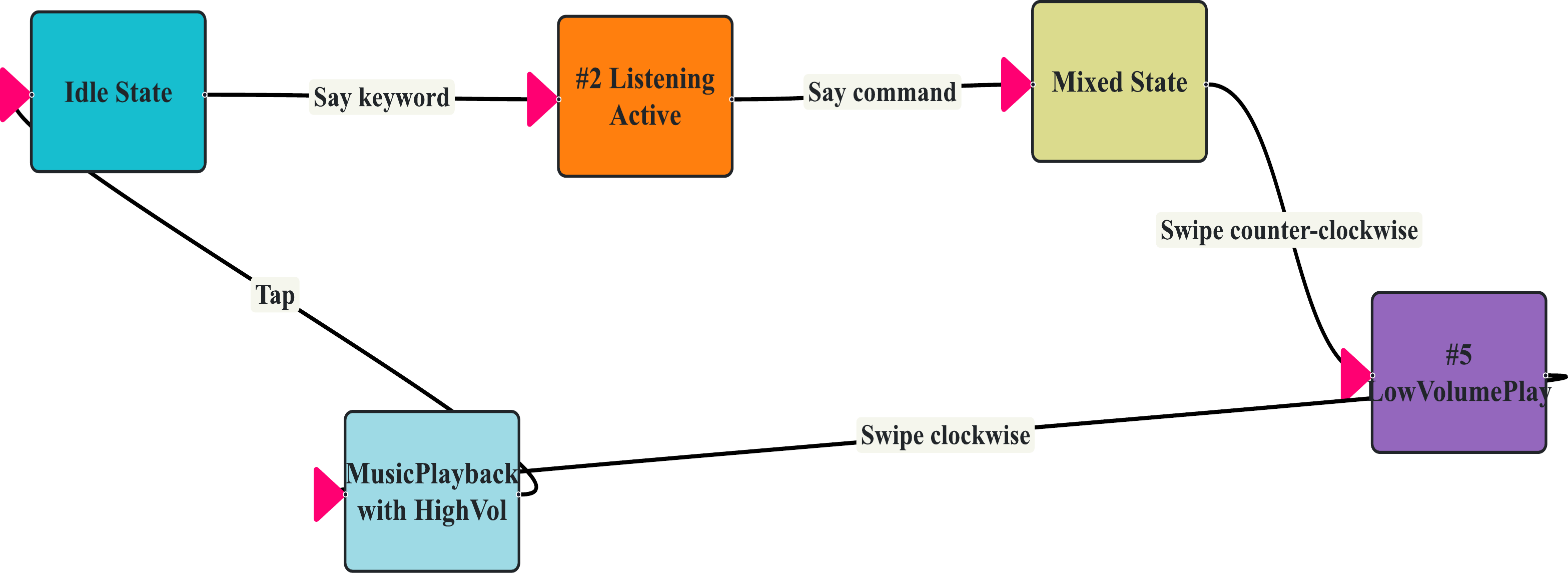}
    \caption{Final FSM from P6.}
    \label{fig:fsm_result:2}
    \end{subfigure}
    \caption{Users may have different interpretations of the IoT device's internal states. This variation may be partly due to the lasting impact of their initial mental models, which affect the way users revise their understanding~\cite{bibby1996instruction}.}
    \label{fig:fsm_result}
\end{figure*}

\vspace{0.1cm}\noindent\textbf{Cognitive Load}. Table~\ref{tb:quanti_2} shows the statistical results of the participants probing the Google Home smart speaker with the metrics from the NASA task load index on a scale of 1 to 21, where a lower score indicates a lower cognitive load). This index assesses the overall workload of a participant by measuring performance across six dimensions: (1) mental demand, (2) physical demand, (3) temporal demand, (4) performance, (5) effort, and (6) frustration. The average, median, and standard deviation value across these six dimensions is around 4.12, 3.83, and 2.87 respectively. As we can see, the task load measured by these metrics is quite low due to the friendly GUI design, indicating \sysname's capability of helping participants probe the IoT device's internal states with low cognitive load.

\subsubsection{Qualitative Feedback}
To answer RQ3, we utilized recordings and transcripts of the interviews, conducted inductive thematic analysis~\cite{thematic_analysis}, and held meetings to review the analysis process and discuss the findings. 


\begin{itemize}
    \item \textbf{Transparency.} \sysname's ability to enhance the transparency of IoT devices, especially the efficiency of probing and verifying the black-box IoT device's states was acknowledged by the participants. One of the key aspects they appreciated was the generated FSM, which made the understanding of the device's functioning more accessible. As P9 mentioned, "\textit{I have used the Google Home smart speaker before, but I never got to know what it actually does inside, its operations are often opaque. This system allows me to understand the device's functioning through the finite state machine}". The transparency is further enhanced by the use of transition events. P2 noted the utility of this feature, saying, "\textit{With the transition event, I can easily recall and track the interactions with the device. It also gives me a clearer understanding of how the device responds to different inputs, allowing me to establish a meaningful connection between different internal states of the IoT device}". Another important design is the data representation. As P5 expressed, \textit{"The abundance of data makes me feel a bit overwhelmed at the beginning. Nevertheless, being able to see the low-level data and my annotations side by side truly enhances my comprehension of the IoT device}". P1 also commented on its effectiveness, stating, "\textit{The scatter plot clearly illustrates the low-level data with its interactive attributes and color encoding, which enables me to probe the internal states of the IoT device in a more intuitive way. To be more specific, the interactive features provide me with rapid feedback dynamically during collaging, and the color encoding enhances the clarity of data representation, making it easier to identify and interpret different states}".


    \item \textbf{Learnability.} All participants discussed the learnability of \sysname. Some people found this system, especially the friendly graphical user interface, easy to learn and efficient to probe the IoT device's state. P9 said, "\textit{The system is new to me, but the workflow is quite reasonable and intuitive}". Conversely, some participants shared a different opinion that "\textit{It took me some time to realize what I should do}" (P8). This result echos the fact that only three out of the ten participants had experience working with IoT devices. 
    
    \item \textbf{Usability.} When discussing the usability of \sysname, the design of the user interface received appreciation from participants (P1-6, P9, P10).P6 noted "\textit{UI is nice and user-friendly}", while P7 mentioned, "\textit{the layout and interaction design of the UI make it easy to use}". 
\end{itemize}

\begin{itemize}
    \item \textbf{Diversity of human mental models.} Participants' answers vary when asked about how they annotate states during the exploration stage. Some participants annotated the states based on their understanding and observation. For example, P1 mentioned that "\textit{It’s mainly based on my intuition and understanding}", and P6 noted that "\textit{I annotate them based on my observation about what the device will perform}". However, other participants like P7 relied on different information, saying, "\textit{from the actions I conducted, I can know the state it enters}". This result echoes the situation of semantic redundancy during the exploration stage and implies that the FSM is highly personalized. It also indicates the necessity for a human-in-the-loop design approach, rather than a one-size-fits-all solution.
    \item \textbf{Integration of sensing model and mental model.} Furthermore, to explore how the human-in-the-loop design improves understanding of black-box IoT devices, we asked participants about how they balance the mental model and the sensing model during collaging. A few participants (P1, P3, P4) preferred relying solely on a single model, with statements like "\textit{I collage the states with the same semantics}" from P4, or "\textit{I just collage the states based on the scatter plot}" by P1. However, the majority (P2, P5-10) actively combined both models. For instance, P5 described their approach as "\textit{I make high-level groups based on my understanding, and then refer to the sensing model for verification}". This result underscores the significant roles both the sensing model and the mental model play in the probing process. It also indicates our design objective of incorporating human intelligence alongside sensing techniques within the probing framework.

\end{itemize} 

%% file: ubicomp_parts/p7-related.tex
\section{Related work}
\label{sec:relatedwork}
\vspace{0.1cm}\noindent\textbf{Side-channel sensing}. Many studies have investigated the side-channel information of IoT devices, such as power consumption~\cite{jung2022light}, network traffic~\cite{wan2021iotathena,apthorpe2017smart,huang2020iot,bezawada2018behavioral,oconnor2019homesnitch}, wireless communication~\cite{jin2015corona,zachariah2019browsing}, and acoustic emanations~\cite{asonov2004keyboard,zhuang2009keyboard}. However, these works cannot characterize the fine-grained IoT states. For example, Light auditor~\cite{jung2022light} leverages power consumption measurements to identify the malicious behaviors of exfiltrating information from smart bulbs, while these malicious behaviors are not classified. Network traffic-based IoT state prediction is straightforward, as every IoT device needs to have network access, and the network traffic pattern can be exploited to predict IoT states. For example, IoTAthena~\cite{wan2021iotathena} leverages the raw time-stamped IP packets to predict IoT activities in a coarse-grained manner. Emanations from IoT devices have been exploited to detect the existence of IoT devices (e.g., hidden spy camera detection for human privacy~\cite{liu2023camradar}), while they have never been used for IoT state prediction. Electromagnetic sensing has been widely explored for human-computer interaction~\cite{laput2015sense, langerak2020optimal, langerak2020omni}, which has different intents compared to our EM noise-based IoT state detection. For example, EM-Sense~\cite{laput2015sense} leverages the EM noise generated by everyday electrical and electromechanical objects to achieve touch recognition.

\vspace{0.1cm}\noindent\textbf{IoT program debugging}. Programming language technologies have been extensively employed in the realm of IoT for security and privacy analysis~\cite{adebayo2020debugging}. For instance, TZSlicer~\cite{ye2018tzslicer} introduces a framework that automatically identifies code segments requiring protection. IOTA~\cite{newcomb2017iota}, a core calculus for IoT automation, facilitates the development of conflict detection and provenance in home automation programs. IotSan~\cite{nguyen2018iotsan} is another framework that utilizes model checking to pinpoint undesirable cyber states for IoT devices, offering counter-examples to illustrate the underlying causes. SOTERIA~\cite{celik2018soteria} extracts state models from IoT code, assisting in the detection of security, safety, and functional errors. However, these studies predominantly address IoT device behaviors through a white-box approach, assuming the availability of the device's source code. In contrast, our work focuses on the underexplored area of black-box IoT devices raised in~\cite{huynh2023conversational} and novel device mechanisms, aiming to enhance support for black-box debugging requirements.


\vspace{0.1cm}\noindent\textbf{Tools for IoT privacy.} Considerable efforts have been devoted to creating new tools that strengthen privacy protection in the IoT domain~\cite{zhang2023helping,he2021sok,kuzminykh2021challenges}. For example, Ren et al.\cite{ren2019information} utilized multidimensional analysis to characterize information exposure in IoT devices, emphasizing the importance of privacy preservation. Numerous studies\cite{miettinen2017iot,bao2020iot,ammar2020autonomous, meidan2017profiliot} have categorized IoT devices by analyzing network traffic patterns for authentication purposes. Additionally, Saidi et al.\cite{saidi2020haystack} developed a scalable approach to detecting IoT devices in real-world scenarios using network traffic data. IoT Inspector\cite{huang2020iot}, an open-source tool, facilitates the examination of network traffic for IoT devices in home networks to ensure data privacy. These studies typically examine general IoT privacy on a macro level, focusing on high-level device behavior and the data transmitted. In contrast, our work aims to support privacy enthusiasts at a micro level, enabling them to debug the fine-grained internal state of each individual device.

%% file: ubicomp_parts/p8-conclusion.tex
\section{Conclusion}
\label{sec:cons}
In this paper, we design \sysname, a system that can probe the fine-grained internal states of black box IoT devices using side-channel information such as power consumption, network traffic, and emanations. Furthermore, we design an annotation interface to semantically probe and verify the IoT device's internal states. Our experimental results and user study demonstrate the feasibility of using side-channel information to help users form more accurate mental models. 


%% file: ubicomp_parts/p10-ackknowledgement.tex
\section*{Acknowledgements}
This work is based upon work supported in part by the Office of the Director of
National Intelligence (ODNI), Intelligence Advanced Research Projects Activity
(IARPA), via [2021-2106240007]. The views and conclusions contained herein
are those of the authors and should not be interpreted as necessarily representing
the official policies, either expressed or implied, of ODNI, IARPA, or the U.S.
Government. The U.S. Government is authorized to reproduce and distribute
reprints for governmental purposes notwithstanding any copyright annotation
therein. We also thank Raymond Song, Keith Kwong, and Chan Heng Chan for the early experiments.

%% file: ubicomp_parts/p9-appendix.tex
\appendix

\section{\sysname's GUI Tool}

\subsection{Code for \sysname}
 We plan to release the code of \sysname's design and make it publicly available. 

\subsection{Instruction Table for Google Home Smart Speaker}
\begin{figure}[h]
    \centering
    \includegraphics[width=0.5\linewidth]{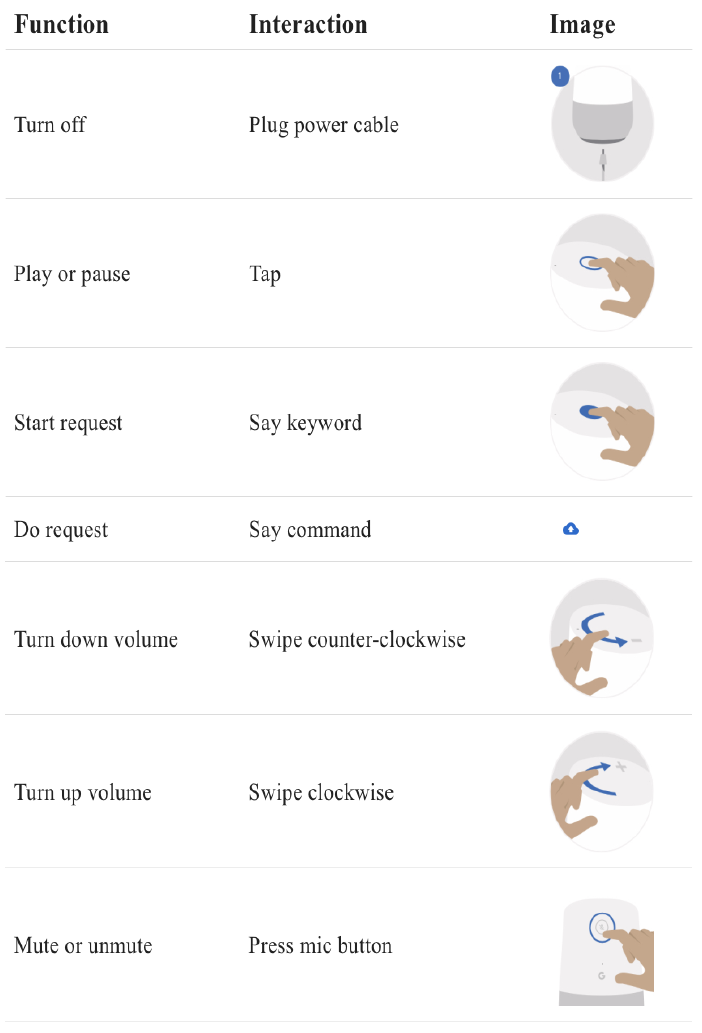}
    \caption{The instruction table shows the users how to interact with the Google Home smart speaker.}
    \label{fig:aiy:fsm}
\end{figure}

As part of \sysname's GUI, we show the instruction table of the Google Home smart speaker to instruct the users to interact with the Google Home smart speaker as shown in Fig.~\ref{fig:aiy:fsm}.

\section{Interview Questions}
\subsection{General Background Questions}
\begin{itemize}
    \item First name [Study use only]
    \item Gender
    \item Education level
    \item Major
\end{itemize}

\subsection{Prior Experience of Using IoT Devices}
\begin{itemize}
    \item Do you have experience in working with IoT devices? If "yes", what is it about?
    \item Do you use any IoT device in your daily life?
    \item Have you used Google Home smart speaker before? If "no", have you used any other smart speakers before?
\end{itemize}

\subsection{Open-Ended Feedback Questions}
\begin{itemize}
    \item How do you annotate states during exploring? What criteria do you use?
    \item What criteria do you use when you collage states?
    \item Do you find the sensing model useful during this process? Why or why not?
        \begin{itemize}
            \item When do you refer to the sensing model?
            \item What type of information does the sensing model provide to you?
        \end{itemize}
    \item Do you find the mental model useful during this process? Why or why not?
        \begin{itemize}
            \item When do you refer to the mental model?
            \item Do you refer to the context information in the UI? If so, when?
            \item What type of information does the mental model provide to you?
        \end{itemize}
    \item Have you encountered any conflict between your mental model and the sensing model?
        \begin{itemize}
            \item What is the conflict? When does it occur?
            \item How do you deal with this conflict?
        \end{itemize}
    \item What do you think of \sysname?
        \begin{itemize}
            \item Do you find it easy to use? Why?
            \item What do you think of this workflow?
            \item Do you think \sysname helps you understand the IoT device? If so, how?
            \item What do you think can be improved in \sysname?
        \end{itemize}
\end{itemize}